\documentclass[12pt]{article}

\usepackage{pdfsync}

\usepackage{amssymb,amsmath,amsfonts,amssymb}%numbysec}
-\marginparwidth \oddsidemargin -9mm \evensidemargin \oddsidemargin
\usepackage{latexsym}
\advance\hoffset by 5mm

\def\@abssec#1{\vspace{.05in}\footnotesize \parindent .2in
{\bf #1. }\ignorespaces}
\newtheorem{theorem}{Theorem}[section]
\newtheorem{thm}[theorem]{Theorem}
\newtheorem{lemma}[theorem]{Lemma}
\newtheorem{proposition}[theorem]{Proposition}

\newtheorem{prop}[theorem]{Proposition}%[section]

\newcommand{\commentout}[1]{{}}

\def \Rm {\mathbb R}

\def \Em {\mathbb E}
\def \Mm {\mathbb M}

\newcommand{\si}{\sigma}

\newcommand{\eps}{\varepsilon}

\newcommand{\pdr}[2]{\dfrac{\partial{#1}}{\partial{#2}}}

\newcommand{\Om}{\Omega}

\newcommand{\farc}{\frac}

\newcommand{\ga}{\gamma}
\newcommand{\bbP}{\mathbb P}

\newcommand{\qed}{$\Box$}

\newcommand{\bd}{d\hspace{-0.9ex}\vphantom{1}^{-}\!}

\newcommand{\al}{\alpha}
\newcommand{\bbR}{\mathbb R}
\newcommand{\bbE}{\mathbb E}
\newcommand{\bbM}{\mathbb M}

\newcommand{\la}{\lambda}
\allowdisplaybreaks \numberwithin{equation}{section}

\newcommand{\E}{\Em}

\author{Thomas Chen\thanks{Department of Mathematics, University of Texas at Austin, 2515 Speedway C1200,
Austin, TX 78712, USA; tc@math.texas.edu}\and Tomasz Komorowski\thanks{Institute of Mathematics,
Maria Curie Sklodowska University,
pl. Marii Curie Sklodowskiej 1,
20-031 Lublin,
Poland;
komorow@hektor.umcs.lublin.pl}
\and Lenya Ryzhik\thanks{Department of Mathematics,
Stanford University,
Stanford, CA 94305, 
USA; ryzhik@stanford.edu }}

\title{The weak coupling limit for the random Schr\"odinger equation: The
 average wave function}
\begin{document}

\maketitle

\begin{abstract}
We consider the Schr\"odinger equation with a time-independent weakly random  potential
of a strength $\eps\ll 1$, with Gaussian  statistics. 
We prove that when the initial condition varies on a scale  much larger than the correlation
length of the potential, the compensated wave function converges to a deterministic limit on the
time scale $t\sim\eps^{-2}$.
This is shown under the sharp assumption that the correlation function $R(x)$ of the random potential
decays slower than $1/|x|^2$, which ensures that the effective potential is finite. When $R(x)$ decays slower
than $1/|x|^2$ we establish an anomalous diffusive behavior for the averaged wave function on a time
scale shorter than $\eps^{-2}$, as long as the initial condition is "sufficiently macroscopic".  
We also consider the kinetic regime  when the initial condition varies
on the same scale as the random potential and obtain the limit of the averaged wave function
for potentials with the correlation functions decaying faster than $1/|x|^2$. We use random potentials of the
Schonberg class which allows us to bypass the oscillatory phase estimates. 
\end{abstract}

\section{Introduction}
 
We consider the large time behavior of the
solutions of the weakly random Schr\"odinger equation
\begin{equation}\label{may2302}
i\pdr{\psi}{t}+\frac{1}{2}\Delta\psi-
\eps V(x)\psi=0,~~t>0,~~x\in\Rm^d.
\end{equation} 
Here, the random potential $V(x)$ is a mean-zero Gaussian statistically
homogeneous random field {over a probability space $(\Om,{\cal V},\bbP)$} with the covariance function $R(x)$:
\begin{equation}\label{may2304}
R(x)=\Em(V(y)V(x+y)).
\end{equation}
We denote by $\bbE$ the expectation with respect to $\bbP$. 
The small parameter $\eps\ll 1$ measures the relative strength of the
random fluctuations. The initial condition for (\ref{may2302}):
\begin{equation}\label{may2304bis}
\psi(0,x)=\psi_0\Big(\farc{x}{l_i}\Big),
\end{equation}
varies on a scale $l_i$. Scale-wise, it is implicitly assumed in (\ref{may2302}) 
that the random potential $V(x)$ varies on a scale $l_c=O(1)$.

The Schr\"odinger equation (\ref{may2302}) preserves the total mass
\[
M(t)=\int_{\Rm^d}| \psi(t,x)|^2dx=M(0),
\]
and the total energy
\[
E(t)=\int_{\Rm^d}\Big[\farc{1}{2}|\nabla\psi(t,x)|^2+
\eps V(x)|\psi(t,x)|^2\Big]dx
=\int_{\Rm^d}|\xi|^2|\hat\psi(t,\xi)|^2\bd \xi+
\eps\int_{\Rm^d}V(x)|\psi(t,x)|^2dx.
\]
Both here and in what follows we denote $
\bd p:=dp/(2\pi)^d$.  We also use the notation
\[
\hat\psi(t,\xi)=\int_{\bbR^d} e^{-i\xi\cdot x}\psi(t,x)dx
\]
for  the Fourier transform of the wave function.

Since the random potential is weak, the preservation of 
the energy, together with the mass conservation, means, 
approximately, that the bulk of the energy
would remain at the frequency scale $|\xi|\sim l_{i}^{-1}$ of the initial condition.

\subsubsection*{The kinetic regime}

This problem
has been extensively studied in the past when the random potential is
rapidly decorrelating and the initial condition varies on the same scale
as the random potential. In other words,  $l_i=1$, or, in the dimensional variables, $l_i=l_c$. This is sometimes known as the  kinetic regime, 
and is particularly interesting
since it leads to a full interaction between the random fluctuations and the
wave function. The first key result here is by Spohn~\cite{Spohn} who considered
the rescaled wave function 
\[
\psi_\eps(t,x)=\psi\Big(\farc{t}{\eps^2},\farc{x}{\eps^2}\Big),
\]
and its Wigner transform
\begin{equation}\label{may2306}
W_\eps(t,x,\xi)=\int_{\bbR^d} e^{i\xi\cdot y}
\psi_{\eps}\Big(t,x-\farc{\eps^2 y}{2}\Big)\psi^*_{\eps}\Big(t,x+\farc{\eps^2 y}{2}\Big)
\bd y.
\end{equation}
Here, $z^*$ denotes the complex conjugate of $z\in\mathbb C$.
The main result of \cite{Spohn} is that, if the correlation function $R(x)$
is smooth and sufficiently rapidly decaying, then 
$\Em(W_\eps(t,x,\xi))$ converges, as~$\eps\to 0$, in the sense of distributions
to the solution~$W(t,x,\xi)$ of the radiative transport equation
\begin{equation}\label{may2308}
\pdr{W(t,x,\xi)}{t}+\xi\cdot\nabla_\xi W(t,x,\xi)=
\int_{\Rm^d}\hat R(\xi-p)\delta\left(\frac{|\xi|^2-|p|^2}{2}\right)(W(t,x,p)-W(t,x,\xi))
\frac{d p}{(2\pi)^{d-1}}.
\end{equation}
Here, $\hat R(p)$ is the power energy spectrum, the
Fourier transform of the correlation function~$R(x)$.

%$\bullet$ {\textcolor{red}{\Large\em I don't follow,
%is $|\psi_\eps(t,x)|^2$ 
%energy after all? Shouldn't we rather talk about the average density?}

It follows that the average density of the solution has the weak limit
\begin{equation}\label{may2310}
\lim_{\eps\to0+}\Em|\psi_\eps(t,x)|^2=\int_{\Rm^d}W(t,x,\xi)d\xi.
\end{equation}
This result has been established in~\cite{Spohn}
in dimensions $d\ge 3$,
on a finite time interval $0\le t\le T$ with~$T$
that is independent of $\eps$ (thus, in the original microscopic variables the 
time interval is~$T/\eps^2$) but that does depend on the correlation function
$R(x)$. The assumption that the power energy spectrum satisfies
$\hat R\in L^1\cap L^\infty$ is 
essential for the estimates in \cite{Spohn}.
   
The kinetic limit has been further studied by L. Erd\"os and H.~T. Yau
in~\cite{Erdos-Yau2}. They have removed the restriction of a finite time
interval convergence, and have shown that 
the kinetic limit holds on any finite time
interval $0\le t\le T$. That is, microscopically, it is valid on any time
interval of the size $O(\eps^{-2})$. The assumptions
on the decay of the correlation function in~\cite{Chen,Erdos-Yau2} are %slightly
more stringent than in \cite{Spohn}. 
See also \cite{bo1,bo2,ch1,lusp,pova} for related results.
Convergence of the expectation
has been strengthened to the~$L^2$-convergence in~\cite{Chen}, 
with some further improvements obtained in~\cite{bu}. 
 
The results of \cite{Erdos-Yau2} were subsequently extended to the analysis of the
diffusive limit in \cite{ESY1,ESY2}. For random Schr\"odinger equations coupled to 
thermal noise, diffusive limits have been studied in \cite{frdrpi,frsch,kasch}.

\subsubsection*{The homogenization regime}

Another regime recently investigated by G. Bal and
N. Zhang in~\cite{ZB-CMS-14,ZB-SD-14} is $l_i=\eps^{-1}l_c$.
That is, the initial condition for (\ref{may2302})
is of the form $\psi(0,x)=\psi_0(\eps x)$. This is an interesting and convenient 
regime as in this case 
the central limit time scale $O(\eps^{-2})$, which comes from the size~$\eps$
of the random potential, matches the 
homogeneous Schr\"odinger time scale $t\sim l_i^2$, which is the time for
the linear "Schr\"odinger phase" $\exp(it|\xi|^2/2)$ to become of the order
$O(1)$ for~$|\xi|\sim l_i^{-1}$. Accordingly, when $l_i=\eps^{-1}$,
a natural object is the rescaled wave function  
\[
\psi_\eps(t,x)=\psi\Big(\farc{t}{\eps^2},\frac{x}{\eps}\Big).
\]
It has been shown 
in~\cite{ZB-SD-14} in $d\ge 3$ that $\psi_\eps(t,x)$
converges in probability, as $\eps\to 0$ to the solution 
of the deterministic homogenized
problem
\begin{equation}\label{may2316}
i\pdr{\psi_*}{t}+\farc 12\Delta\psi_*-R_*\psi_*=0,
\end{equation}
with the initial condition $\psi_*(0,x)=\psi_0(x)$. The effective potential
in (\ref{may2316}) is simply a constant
\begin{equation}\label{may2318}
R_*=\int_{\Rm^d}\farc{\hat R(p)}{|p|^2}\bd p.
\end{equation}
Thus, there is a substantial difference in the behavior of the solutions
when $l_i=1$ and $l_i=\eps^{-1}$. In both cases, the solution is affected in
a non-trivial way by the central limit time scale $t\sim\eps^{-2}$.
However, in the former case, the solution
behaves stochastically on this time scale -- this is the kinetic regime, 
while in the latter it has a deterministic behavior -- this is the homogenization
regime.

A stochastic (non-deterministic) limit for such problems when the Laplacian
is replaced by a higher power $(-\Delta)^m$ with $m>1$ has been investigated 
in~\cite{ZB-CMS-14}, also for rapidly decorrelating potentials.

\subsection*{The main results}

Our main interest in this paper is in the breakdown
of the homogenization regime when the effective potential $R_*$ in 
(\ref{may2318}) is infinite. However, all of the above results have been
established under much more stringent assumptions on the
correlation function than just
\begin{equation}\label{may2320}
R_*<+\infty.
\end{equation}
Hence, we first carry out the analysis of the behavior of the
wave function only assuming 
(\ref{may2320}), to reach the threshold of the validity of the homogenization 
and kinetic regimes. We also obtain some results when $R_*=+\infty$, and
these regimes break down.

\subsubsection*{The correlation function} 

In order to perform the analysis under weaker assumptions on the
decay of the correlation function than  
previously, 
we take a special form of the correlation function.
We shall assume that the power energy spectrum of the
potential~$V(x)$ is  positive definite in the sense
of Schoenberg~\cite{schoenberg}. That is, $\hat R(p)$ is of the form 
\begin{equation}
\label{schoen1abis}
\hat R(p)=\int_1^{+\infty}e^{-\la^2|p|^2/2}s(\lambda)\frac{ d\la}{\la^{\ga}},
\end{equation}
with a positive bounded function $s(\lambda)$ that satisfies:
\[
0<c\le s(\lambda)\le c^{-1}, \quad \forall\la>1,
\]
for some $c>0$. It is easy to see that this assumption is not restrictive
in term of the spatial decay of the correlation function.
The power spectrum has the asymptotics
\begin{equation}\label{jun3002}
\hat R(p) {\sim} |p|^{\ga-1},\quad |p|\ll1,
\end{equation}
which translates into the following spatial decay for the 
correlation function itself:
\begin{equation}\label{may2322}
R(x)\sim\frac{1}{|x|^{d+\gamma-1}},~~~|x|\gg 1.
\end{equation}
Since the potential $V(x)$ is a Gaussian random field, 
we have $\Em(V^2(x))<+\infty$, thus 
$\hat R\in L^1(\bbR^d)$, and (\ref{jun3002}) 
implies that $\ga>1-d$. On the other hand, choosing 
various~$\gamma>1-d$ in~(\ref{schoen1abis}), 
we may achieve an arbitrarily slow or fast decay of the correlation function in (\ref{may2322}).
%We will require that
%\begin{equation}\label{finitebis}
%R_*:=\int_{\bbR^d}\frac{\hat R(p)\bd p}{|p|^2}<+\infty.
%\end{equation}
A straightforward computation shows that the effective potential in
(\ref{may2318}) is
given~by
\begin{eqnarray}\label{may2324}
&&
R_*%=\int_{\Rm^d}\frac{\hat R(p)}{|p|^2}\farc{dp}{(2\pi)^d}
%=\frac{\om_{d-1}}{(2\pi)^d}\int_1^\infty\frac{
%s(\lambda)  d\la}{\la^{\ga}}\int_0^\infty e^{-\la^2r^2/2} r^{d-3} dr
%=\frac{\om_{d-1}}{(2\pi)^d}\left(\int_1^\infty
%\frac{s(\lambda) d\la}{\la^{\ga+d-2}}\right)
%\left(\int_0^\infty e^{-r^2/2} r^{d-3} dr\right)\nonumber\\
%&&
=
\farc{1}{(2\pi)^{d/2}(d-2)}\int_1^\infty\frac{
s(\lambda) d\la}{\la^{\ga+d-2}}.
%~~C_d=\frac{2^{d/2-2}\om_{d-1}}{(2\pi)^d}\Gamma\left(\frac{d}{2}-1\right).
\end{eqnarray}
%Here $\Gamma(\cdot)$ is the Euler gamma function, $\om_{d-1}$
%$$
%\om_{d-1}=\frac{2\pi^{d/2}}{\Gamma(d/2)}
%$$
%is the surface area of the unit sphere in $\mathbb R^d$, and
%\begin{equation}
%\label{s-star}
%S_*:=\int_1^\infty\frac{
%s(\lambda) d\la}{\la^{\ga+d-2}}.
%\end{equation}
Thus,  the effective potential is finite: $R_*<+\infty$, provided that
$\gamma>3-d$, or, in terms of the decay of the correlation function, that
\begin{equation}\label{may2326}
R(x)\sim\farc{1}{|x|^m},~~|x|\gg 1,
\end{equation}
with  {some} $m>2$. This assumption does not depend on the spatial dimension.

We should mention that,  similarly to the very technical proofs 
in~\cite{Erdos-Yau2,Spohn}, our strategy relies on the Duhamel expansion of the
solutions of the Schr\"odinger equation, and summation of the
Feynman diagram expansions that come up after we evaluate various expectations.
Typically, this requires intricate oscillatory phase estimates. One
contribution of this paper is a simple observation that
for the random potentials that are in the
Schoenberg class, we may bypass the oscillatory phase arguments. 
Instead, one needs to estimate some quite
explicit determinants,  simplifying substantially
the analysis and allowing it to go a bit deeper.
%We conclude therefore that  thus also 
% \eqref{finite} holds, provided that the above condition holds and $d\ge3$. After a simple calculation
% we obtain
% \begin{equation}
% \label{R-star}
% R_*=\frac{S_*}{(2\pi)^{d/2}d}.
% \end{equation} 
%Moreover,   we have the asymptotics
%$$
%\hat R(p)=O( |p|^{\ga-1}),\quad |p|\ll1,
%$$
%and
%$$
%\hat R(p)\ll e^{-|p|^2},\quad |p|\gg1.
%$$
%

\subsubsection*{The microscopic initial conditions}

As has been observed in~\cite{bkr}, it is convenient to take out
the rapidly growing phase $\exp(it|\xi|^2t/2)$
and consider
the compensated wave function     
\begin{equation}\label{may2330}
\hat\zeta(t,\xi):=\hat\psi(t,\xi)e^{i|\xi|^2t/2}.
\end{equation}
As long as $R_*<+\infty$, we will be interested in the central limit time scales
$t\sim\eps^{-2}$, and, accordingly, define
\begin{equation}\label{may2332}
\hat\zeta_\eps(t,\xi):=\hat\psi\left(\farc{t}{\eps^2},\xi\right)e^{i|\xi|^2t/(2\eps^2)}.
\end{equation}
We first consider the initial condition for (\ref{may2302}) with $l_i=1$ -- this
is the kinetic regime considered in~\cite{Erdos-Yau2,Spohn}, or, alternatively,
the case of microscopic initial conditions. 
Let us define
\begin{equation}
\label{may2332bis}
r(\xi)=
{\frac{i}{2(2\pi)^{d/2}}}\int_1^\infty\frac{ s(\lambda)\kappa(\xi,\la) d\la}{\la^{\ga+d-2}},
\end{equation}
with
\begin{equation}
\label{kappa}
\kappa(\xi,\la):=\int_0^{+\infty}\frac{1}{(1+i\tau)^{d/2}}\exp
\left\{-\frac{(\la|\xi|\tau)^2}{2(1+i\tau)}\right\}
d\tau.
\end{equation}
It is straightforward to verify that the real part of $r(\xi)$ is the
total scattering cross-section in the radiative transport equation (\ref{may2308}).
%{\bf CHECK THE $2\pi$ factors and such.}
\begin{thm}
\label{thm-may2302}
Suppose that $d\ge 3$ and $\ga>3-d$. Let $\psi(t,x)$ be the solution
of (\ref{may2302}) with the initial 
condition $\psi(0,x)=\psi_0(x)\in {\cal S}(\Rm^d)$. 
Then, there exists $t_0>0$ such that
for all $t\in[0,t_0]$ and $\xi\in\bbR^d$, we have
\begin{equation}
\label{232611}
\lim_{\eps\to0+} {\mathbb E}\hat\zeta_\eps(t,\xi)=\hat \psi_0(\xi)\exp\left\{ir(\xi)t
\right\}.
\end{equation}
\end{thm}
This result is not surprising -- the imaginary part of $r(\xi)$ agrees with the
total scattering cross-section for the kinetic equation obtained  
in~\cite{Erdos-Yau2,Spohn}. However, in terms of the assumptions on the
correlation function $R(x)$, it holds up to the threshold $\gamma=3-d$,
when the effective potential becomes infinite. 
On the other hand, as 
$r(0)=-R_*$, the result cannot be true (at least for~$\xi=0$) when~$\gamma\le 3-d$, since in that
case $R_*=+\infty$.
Let us add that the assumption that $\psi_0$
is in the Schwartz class can be easily improved but it is not the focus of this
paper.

Preliminary computations indicate that the convergence of the expectation in 
Theorem~\ref{thm-may2302} may be bootstrapped to the convergence of the second
moment, recovering, in addition, 
the kinetic limit of~\cite{Erdos-Yau2,Spohn}. One may also combine the
techniques of the present paper with the strategy of~\cite{Erdos-Yau2} to
extend the result to all times $t_0>0$. However,
to keep the paper relatively
short, we postpone these directions for a future investigation.

\subsubsection*{The macroscopic initial conditions: homogenization}

Unlike the microscopic initial conditions with the width $l_i=1$,
the macroscopic initial conditions have the initial pulse
width 
$
l_i=\eps^{-\beta}\gg l_c=1,
$
with
some~$\beta>0$. Recall that
the special case~$\beta=1$ has been considered in~\cite{ZB-SD-14} for very
rapidly decorrelating potentials.
In other words,   the initial condition for (\ref{may2302}) is of the form
\begin{equation}
\label{may2336}
\psi(0,x)=\eps^{d\beta/2}\psi_0(\eps^{\beta} x),
\end{equation}
with $\psi_0\in{\cal S}(\Rm^d)$.
The pre-factor in (\ref{may2336}) is introduced simply to keep 
the $L^2$ norm of the solution
to be of order $O(1)$. Its Fourier transform is
\[
\hat\psi(0,\xi)=\eps^{-d\beta/2}\hat\psi_0(\eps^{-\beta} \xi).
\]
To take into account the lower frequencies of the macroscopic
initial conditions, and the 
aforementioned fact
that the bulk of the energy is expected to stay at the original frequency, 
the compensated wave function $\hat\zeta_\eps(t,\xi)$ on the time
scales $t\sim\eps^{-2}$ is now defined
as
\begin{equation}\label{def-zeta-eps1}
\hat\zeta_\eps(t,\xi):=\eps^{d\beta/2}\hat\psi
\left(\frac{t}{\eps^2},\eps^{\beta}\xi\right)e^{i\eps^{2(\beta-1)}|\xi|^2t/2}.
\end{equation}
This allows us to track frequencies of the order $O(\eps^\beta)$, present in the 
initial condition, at the central limit time scale $t\sim O(\eps^{-2})$.
\begin{thm}
\label{main1}
Suppose that $d\ge 3$ and $\ga>3-d$ and the initial condition
for \eqref{may2302} is of the form
\eqref{may2336}. Then,  there exists $t_0>0$ such that
for all~$t\in[0,t_0]$ we have
\begin{equation}
\label{232611a}
\lim_{\eps\to0+}\bbE\|\hat\zeta_\eps(t,\cdot)-\bar\zeta(t,\cdot)\|_{L^2(\bbR^d)}=0,
\end{equation}
with
\begin{equation}
\label{may2338}
\bar\zeta(t,\xi):=\hat \psi_0(\xi)\exp\left\{-i{R_*t}\right\},
\end{equation} 
with the effective potential $R_*$ given by \eqref{may2324}.
\end{thm}
%{\bf THERE WAS $\kappa(0)$ before which was equal to a multiple of $R_*$ not $R_*$ 
%which is probably a matter of agreeing of constants.}
Thus, the homogenization result of~\cite{ZB-SD-14} holds not just for the 
somewhat artificial choice $l_i=\eps^{-1}$ but essentially for all $l_i\gg 1$.
The threshold to the stochastic behavior is exactly at $l_i=l_c$.
The result of Theorem~\ref{main1} is sharp in terms of the
assumptions on the correlation function~-- its conclusion
holds for all random potentials such that the effective potential $R_*<+\infty$.

\subsubsection*{The macroscopic initial data: superdiffusive behavior} 
 
Next, we consider slowly decorrelating 
random potentials, with the power spectrum of the form~(\ref{schoen1abis}), and $\gamma<3-d$, so that $R(x)$ decays
at a rate slower than $1/|x|^2$ as $|x|\to+\infty$ -- see (\ref{may2322}). Then, 
the effective potential is infinite:
$R_*=+\infty$, and the homogenization limit can not hold. 
We assume that the initial condition is macroscopic: $l_i=\eps^{-\beta}$
with some~$\beta>0$. Typically, in such situations one expects a non-trivial
effect of the random fluctuations to be seen on a time scale 
$t\sim \eps^{-2\alpha}$ with some $\alpha\in(0,1)$, rather than for $t\sim\eps^{-2}$.
Accordingly, 
the compensated wave function $\hat\zeta_\eps(t,\xi)$ is now defined
as 
\begin{equation}\label{def-zeta-eps2}
\hat\zeta_\eps(t,\xi):=\eps^{d\beta/2}\hat\psi^{(\eps)}\left(\frac{t}{\eps^{2\al}},\eps^{\beta}\xi\right)e^{i\eps^{2(\beta-\al)}|\xi|^2t/2}.
\end{equation}
We denote the standard Brownian motion by $B_t$ and the expectation with 
respect to it by $\bbM$.
\begin{thm}
\label{main1a}
Suppose that $d\ge 3$ and $\beta>\al$, where
\begin{equation}
\label{010912}
\al:=\frac{2}{5-\ga-d}
\end{equation}
and $\ga\in(1-d,3-d)$. Then,  there exists $t_0>0$ such that
\begin{equation}
\label{011012}
\bar \zeta(t,\xi):=\lim_{\eps\to0+}\Em \hat\zeta_\eps(t,\xi)=
\hat \psi_0(\xi)\bbM \exp\left\{-\frac{ {\mathfrak R}  t^{H}}{4}\int_0^1\int_0^1
 |B_s-B_{s'}|^{1-\ga-d}dsds'\right\},
\end{equation}
for all $t\in[0,t_0]$ with
%\begin{equation}\label{aug2710a}
%\bar\zeta(t,\xi)=\hat \psi_0(\xi)\bbM \exp\left\{-\frac{ {\mathfrak R}  t^{H}}{4}\int_0^1\int_0^1
% |B_s-B_{s'}|^{1-\ga-d}dsds'\right\},
%\end{equation}
%where 
$H=1/\al$, and a constant $\mathfrak R$ whose real part $\hbox{\rm Re}\,\mathfrak R>0$.
%\begin{equation}
%\label{frakR}
% {\mathfrak R}:=\int_0^{+\infty}u^{(\ga+d-3)/2}e^{i u }du.
%\end{equation}
\end{thm}
The assumption $\beta>\alpha$ informally means that the initial condition is "very macroscopic".
A short computation, starting from (\ref{011012}) shows
that there exists $C>0$ such that
\begin{equation}\label{may2334}
 |\bar\zeta(t,\xi)|\le C\exp\left\{- t^{H_*}/C\right\}, \quad (t,\xi)\in [0,+\infty)\times \bbR^d,
\end{equation}
with
\begin{equation}\label{may2336bis}
H_*=\frac{5-\ga-d}{\ga+d+1}.
\end{equation}
Note that the randomization time $t^{-2\alpha}$ does not depend on $\beta$,
as long as $\beta>\alpha$. Informally, this means that solutions
with all sufficiently slowly varying initial conditions are randomized
at the same time scale $t^{-2\alpha}$. We expect that solutions with the "less macroscopic"
initial conditions varying on a scale $l_i=\eps^{-\beta}$ with $\beta\in(0,\alpha)$
are randomized on time scales that depend on $\beta$, but leave this issue for
a further investigation.

The paper is organized as follows: in Section \ref{sec:resum} 
we derive a representation of $\bbE\hat\zeta_\eps(t,\xi)$ in terms 
of the average of the expectation of the exponential of some functional over 
Brownian paths, see Proposition  \ref{prop010811}. Section \ref{sec3} 
is devoted to the presentation of the proofs of Theorems~\ref{thm-may2302} and \ref{main1}. 
Finally in Section \ref{sec5.3.2} we give the proof of Theorem \ref{main1a},
and the short Section~\ref{sec:aux} contains the proof of a standard auxiliary result.
\\

{\bf Acknowledgments.}  TC was supported by the NSF Career grant DMS-1151414,
TK by the
Polish National Science Center grant DEC-2012/07/B/ST1/03320, and LR  
by an AFOSR NSSEFF Fellowship and NSF grant DMS-1311903.

\section{A Brownian 
formula for  the averaged wave function} \label{sec:resum}

In this section, we consider the Schr\"odinger equation, without any 
assumption on the smallness of the random potential
\begin{eqnarray}\label{schr1}
&&i\pdr{\psi}{t}+\frac{1}{2}\Delta\psi-
V(x)\psi=0,\\
&&\psi(0,x)=\psi_0(x).\nonumber
\end{eqnarray}
We assume that $\psi_0\in {\cal S}(\bbR^d)$
and $V(x)$ is a Gaussian, stationary random field with 
continuous realizations defined over 
a probability space $(\Omega,{\cal F},\mathbb P)$.
The complex valued spectral measure~$\hat V(dp)$
corresponding to the field 
\[
V(x)=\frac{1}{(2\pi)^d}\int_{\Rm^d} e^{ip\cdot x}\hat V(dp),
\]
has the covariance
\[
\bbE \left[\hat V(dp)\hat V^*(dq)\right]=(2\pi)^d\delta(p-q)\hat R(p)dpdq,\quad p,q\in\bbR^d,
\]
with a non-negative function $\hat R\in L^1(\bbR^d)$.
As $V(x)$ is real valued and $\hat R(p)\ge 0$ for all
$p\in\Rm^d$, 
we have~$\hat R(-p)=\hat R(p)$, for all
$p\in\bbR^d$. We do not assume in this section that $\hat R(p)$
has the Schonberg form (\ref{schoen1abis}) -- this will be done
starting with Section~\ref{sec3} onwards.

%The compensated wave function is defined as
%\begin{equation}\label{may2340}
%\hat\zeta(t,\xi):=\hat\psi(t,\xi)e^{i|\xi|^2t/2}.
%\end{equation}
The goal of this section is to obtain a convenient
representation for the average compensated wave function
\begin{equation}\label{may2342}
\bar\zeta(t,\xi)=\bbE\hat\zeta(t,\xi),
\end{equation} 
with $\hat\zeta(t,\xi)$ defined in (\ref{may2330}).
Let us introduce
\begin{equation}
\label{V}
E(t,z,p,\xi):= \exp \left\{z(\sqrt{i}B_t+t\xi)
  \cdot p\right\} ,\quad (t,z,p,\xi)\in[0,+\infty)\times\mathbb C\times \bbR^{2d},
\end{equation}
where $B_t$ is a $d$-dimensional standard Brownian motion
over a probability space $(\Sigma,{\cal A},\mathbb Q)$. 
We denote by
$\bbM$  the
expectation w.r.t. $\mathbb Q$, and set
\begin{equation}
\label{c-n}
c_n(t,\xi):=\mathbb M\left\{\int_0^tds\int_0^t ds'\int_{\bbR^d} 
E(s,p,i;\xi)E(s',p,-i;\xi)\hat
R(p)\bd p\right\}^n.
\end{equation}
The following will be the starting point for our analysis of the asymptotic limits.
\begin{prop}
\label{prop010811} 
We have
\begin{equation}\label{aug2710}
\bar\zeta(t,\xi)=\hat\psi_0(\xi)\sum_{n=0}^\infty \frac{(-1)^n}{(2n)!!}c_n(t,\xi),\quad t\ge0,\,\xi\in\bbR^d,
\end{equation}
or equivalently
\begin{equation}
\label{c-n1}
\bar\zeta(t,\xi)=\hat\psi_0(\xi)\mathbb M\bbE\left\{\exp\left\{ {\frac{i}{(2\pi)^{d}}}\int_{\bbR^d}\hat V(dp)\int_0^tE(s,i,p,\xi)ds\right\}\right\}.
\end{equation}
\end{prop}
The rest of this section contains the proof of this proposition.

\subsection{The Duhamel expansion}

We re-write the Schr\"odinger equation
(\ref{schr1}) as an integral in time equation
\begin{eqnarray}\label{may2346}
&&\hat\psi(t,\xi)
=\hat\psi_0(\xi)e^{-i|\xi|^2t/2} +
\frac{1}{i}
\int_0^t\int_{\bbR^d}  \farc{\hat V(dp_1)}{(2\pi)^d}
\hat\psi(s_1,\xi-{p_1})
e^{-i|\xi|^2(t-s_1)/2}ds_1.
\end{eqnarray}
%Here, $\hat V(dp)$ is the complex valued spectral measure corresponding to the field $V(x)$. We assume that its structure measure is of the form
%$$
%\bbE \left[\hat V(dp)\hat V^*(dq)\right]=(2\pi)^d\delta(p-q)\hat R(p)dpdq,
%$$
%where $\bbE$ is the expectation corresponding to $\mathbb P$ and
% \begin{equation}
% \label{010212}
% \hat R\in L^1(\bbR^d)\quad \mbox{ is non-negative.}
% \end{equation}
%  Since we require the field $V(x)$ to be real valued we have $\hat R(-p)=\hat R(p)$, $p\in\bbR^d$. We specify our further  assumptions on the spectrum of $V(x)$ in
%Section \ref{sec2}.
%Define, the compensated wave function $\hat\zeta(t,\xi)$ as, see \cite{bkr} 
%\begin{equation}\label{def-zeta}
%\hat\zeta(t,\xi):=\hat\psi(t,\xi)e^{i|\xi|^2t/2}.
%\end{equation}
%We shall use the following notation: for $a=(a_1,\ldots,a_d)\in\mathbb
%C$ we let
%$$
%a^2:=\sum_{j=1}^da_j^2,\quad |a|^2:=\sum_{j=1}^d|a_j|^2.
%$$
The compensated wave function $\hat\zeta(t,\xi)$
satisfies 
\begin{equation}\label{zeta-eq1}
\hat\zeta(t,\xi)=\hat\psi_0(\xi)  +
\frac{1}{i}
\int_0^t\int_{\bbR^d}   \farc{\hat V(dp_1)}{(2\pi)^d}
\hat\zeta(s_1,\xi-{p_1})\exp\big\{
{i(|\xi|^2-|\xi-{p_1}|^2)\farc{s_1}{2}}\big\}ds_1.
\end{equation}
Iterating (\ref{zeta-eq1}), we get 
an infinite series expansion for $\hat\zeta(t,\xi)$:
\begin{equation}\label{zeta-expand}
\hat\zeta(t,\xi)=\sum_{n=0}^\infty{\hat\zeta_n}(t,\xi),
\end{equation}
with $\hat\zeta_0(t,\xi)=\hat\psi_0(\xi)$, and the rest of the
individual terms of the form
\begin{equation}\label{zeta-n}
\hat\zeta_n(t,\xi) =\left[\frac{1}{i (2\pi)^{d}}\right]^n
\int_{\Delta_n(t)}ds_{1,n}\int_{\bbR^{dn}} \hat V(dp_1)\dots\hat V( dp_n)
\hat\psi_0\left(\xi-\sum_{j=1}^np_j\right) e^{iG_n}, 
\end{equation}
with the phase
\begin{equation}\label{G-n}
G_n=G_n(s_{1,n},p_{1,n})=\sum_{k=1}^n\left(\left|\xi-\sum_{j=1}^{k-1}p_j\right|^2
-\left|\xi-\sum_{j=1}^{k}p_j\right|^2\right)\frac{s_{n-k}}{2}.
%=A_n(\bs^{(n)},\bp^{(n)})-B_n(\bs^{(n)},\bp^{(n)}). 
\end{equation}
Here, we denote $p_{1,n}:=(p_1,\dots,p_n)\in\Rm^{nd}$, and
$s_{1,n}:=(s_1,\dots,s_n)\in\Rm^n$,  so that
$ds_{1,n}:=ds_1ds_2\dots ds_n$.
We have also denoted in (\ref{zeta-n}) by  $\Delta_n(t)$   the time simplex 
\[
\Delta_n(t)=\{(s_1,s_2,\dots, s_n):~~0\le s_1\le s_1\le\dots\le s_n\le t\}.
\]
The next standard proposition shows that we may employ term-wise
expectation. Its proof is standard, for the convenience
of the reader, we present it in   Section \ref{sec:aux}.
\begin{prop}\label{prop-ser0}
(i) The series \eqref{zeta-expand} for the function $\hat\zeta(t,\xi)$
converges almost surely for any initial data $\psi_0\in {\cal S}(\Rm^d)$.
\\
(ii) For each $(t,\xi)\in\Rm^{1+d}$ fixed, we have
\begin{equation}\label{exp-series}
\E\hat\zeta(t,\xi)=\sum_{n=0}^\infty\E\hat\zeta_{2n}(t,\xi).
\end{equation}
(iii) Moreover, we have
\begin{equation}
\label{011708}
\E\|\hat\zeta(t,\cdot)\|_{L^2(\bbR^d)}^2=\|\hat\psi_0\|_{L^2(\bbR^d)}^2,\quad t\ge0.
\end{equation}
\end{prop}

Note that when we introduce various scaling parameters, this proposition
will not guarantee that the convergence
will be
uniform in those parameters. Thus, while this proposition allows us to
interchange the expectation and the summation of the series, it will
not allow us to pass to the limit in the individual terms of the
series when we will consider appropriate asymptotic limits. 
 
Using Proposition~\ref{prop-ser0}, and the rule of the expectation for
the product of $2n$ mean zero Gaussian random variables, we may re-write the expectation
%\begin{equation}
%\label{101508}
%\bar \zeta(t,\xi):= \mathbb E \zeta(t,\xi)
%\end{equation}
as a series
\begin{equation}
\label{111508}
\bar \zeta(t,\xi):= \mathbb E \zeta(t,\xi)
= \left\{1+\sum_{n=1}^{+\infty}\sum_{{\cal F}\in{\mathfrak F}_{2n}}I({\cal F})(t)\right\}\hat\psi_0(\xi).
\end{equation}
Here, ${\mathfrak F}_{2n}$ is the collection of all
pairings $\cal F$ formed over the set $\mathbb Z_{2n}:=\{1,\ldots,2n\}$.  
The individual terms for each pairing have the form
\begin{equation}\label{zeta-n-cb}
I({\cal F}):=(-1)^n
\int_{\bbR^{2dn}} 
\prod_{(k,\ell)\in{\cal F}} \delta(p_k+p_{\ell})
\hat R(p_k) \bd p_{1,2n}
\int_{\Delta_{2n}(t)}
e^{iG_{2n}}ds_{1,2n},
\end{equation}
where $\bd p_{1,n}=\bd p_{1}\ldots \bd p_{n}$,   
%with
%  the convention
%$\bd p:=(2\pi)^{-d}dp$
and
\begin{eqnarray}\label{G-nbis4bis}
&&G_{2n}(s_{1,2n},p_{1,2n}) =\sum_{m=1}^{2n}(s_{2n-m+1}-s_{2n-m}) \Big( \xi\cdot
\sum_{j=1}^mp_j
-\farc{1}{2}\Big|\sum_{j=1}^mp_j\Big|^2\Big) ,
\end{eqnarray}
where  $s_{0}:=0$.

\subsection{The proof of Proposition~\ref{prop010811}}\label{sec2}

We now prove Proposition~\ref{prop010811} via
an alternative representation for the series
(\ref{111508}) for $\bar\zeta(t,\xi)$. 
%Let us define  
%\begin{equation}
%\label{V}
%E(s,p,z;\xi):= \exp \left\{z(\sqrt{i}B(s)+s\xi)
%  \cdot p\right\} ,\quad z\in\mathbb C,
%\end{equation}
%where $(B(t))$ is a $d$-dimensional standard Brownian motion.
%over a probability space $(\Sigma,{\cal A},\mathbb Q)$. Denote by
%$\bbM$  the
%expectation w.r.t. $\mathbb Q$. Define also
%\begin{equation}
%\label{c-n}
%c_n(t,\xi):=\mathbb M\left\{\int_0^tds\int_0^t ds'\int_{\bbR^d} 
%E(s,i,p;\xi)E(s',-i,p;\xi)\hat
%R(p)\bd p\right\}^n.
%\end{equation}
%\begin{prop}
%\label{prop010811} 
%\textcolor{red}{Assume that $\hat R(p)$ satisfies \eqref{010212}.}
%%is non-negative and belongs
%%to $L^1(\bbR^d)$, then 
%Then,
%\begin{equation}\label{aug2710}
%\bar\zeta(t,\xi)=\hat\psi_0(\xi)\sum_{n=0}^\infty \frac{(-1)^n}{(2n)!!}c_n(t,\xi),\quad t\ge0,\,\xi\in\bbR^d,
%\end{equation}
%or equivalently
%\begin{equation}
%\label{c-n1}
%\bar\zeta(t,\xi)=\hat\psi_0(\xi)\mathbb M\bbE\left\{\exp\left\{i\int_{\bbR^d}\hat V(dp)\int_0^tE(s,i,p;\xi)ds\right\}\right\}
%\end{equation}
%\end{prop}
%{\bf Proof.} 
 % and
% \begin{equation}
% \label{R1}
% \hat R(p)\le Ce^{-|p|^2/C},\quad \forall\,|p|\ge 1,
% \end{equation}
% for some $C>0$. 
Identity  (\ref{aug2710}) can be restated as
\begin{equation}
\label{010811}
\sum_{{\cal F}\in{\mathfrak F}_{2n}}I({\cal F})(t)
=\frac{(-1)^nc_n(t,\xi)}{(2n)!!},\quad \forall\, n\ge1.
\end{equation}
We first find an expression for the left side of (\ref{010811}).
Given a pairing ${\cal F}$, and $m\in\{1,\ldots,2n\}$, let us denote by
$A_m({\cal F})$ the set of all left
vertices~$\ell\le m$ such that the corresponding right vertex
$r$ satisfies~$r>m$, that is, the edge $(lr)$ crosses over~$m$,
with the convention $A_0({\cal F}):=\emptyset$.
Sometimes,   when the pairing is obvious from the context, we simply 
write $A_m$. 
We also denote by ${\cal L}({\cal F})$ the set of
all left vertices of ${\cal F}$.
For a given pairing~${\cal F}$, we have
\[
\sum_{j=1}^kp_j=\sum_{j\in A_k}p_j, 
\]
a.e. in the measure
$$
\prod_{(k,\ell)\in{\cal F}} \delta(p_k+p_{\ell})
\hat R(p_k) \bd p_{1,2n}.
$$
The phase $G_{2n}$ has, therefore, the form (note that the set~$A_{2n}$ is empty)
\begin{eqnarray}\label{G-nbis4}
&&G_{2n}(s_{1,2n},p_{1,2n}) =\sum_{m=1}^{2n-1}(s_{2n-m+1}-s_{2n-m})  \Big( \xi\cdot \sum_{j\in
  A_{m}}p_j-\farc{1}{2}\Big|\sum_{j\in
  A_{m}}p_j\Big|^2\Big).
\end{eqnarray}
This leads to the expression
\begin{eqnarray}
\label{021802abbb}
&&
I({\cal F})(t)=(-1 )^{n}\int_{{\Delta}_{2n}(t)}ds_{1,2n} 
\int_{\bbR^{nd}} \Big[\prod_{\ell\in {\cal L}({\cal F})}\hat R(p_\ell)\Big]\\
&&\times
\exp\left\{\sum_{m=1}^{2n-1}(s_{2n-m+1}-s_{2n-m})\left(-\frac{i}{2}
\Big|\sum_{j\in A_{m}({\cal F})}p_j \Big|^2+i \xi\cdot
\sum_{j\in A_{m}({\cal F})}p_j \right)\right\} 
 \prod_{\ell\in {\cal L}({\cal F})} \bd p_{\ell}.\nonumber
\end{eqnarray}

Next, we re-write $c_n$ defined by (\ref{c-n}), to make it clear that
(\ref{010811}) holds.
To abbreviate somewhat the notation, we will denote $E(s,i,p,\xi)$ by
$E(s,p)$  (see (\ref{V})). 
We can re-write~$c_n(t,\xi)$, see \eqref{c-n}, as
\begin{equation}
\label{070109}
c_n=\mathbb
M\left\{\int_0^t\ldots \int_0^t ds_{1,2n}
  \int_{\bbR^{2nd}} \bd p_{1,2n}\prod_{j=1}^n\hat R(p_{2j-1})\delta(p_{2j-1}+p_{2j})
\prod_{j=1}^{2n}E(s_j,p_j)\right\} .
  \end{equation}
For each $(s_1,s_2,\dots,s_{2n})$, we re-order the times $s_j$ in the
increasing order, and re-label the indices $j$ accordingly, so that
%
%\textcolor{red}{\bf DOTAD DOTAD DOTAD DOTAD DOTAD DOTAD }
%
$$
c_n=\sum_{\si}\int\limits_{\Delta_{2n}(t)} ds_ {1,2n}
\int\limits_{\bbR^{2nd}} \bd p_{1,2n}\prod_{j=1}^n\Big[\hat
R(p_{\si(2j-1)})\delta(p_{\si(2j-1)}+p_{\si(2j)})\Big]\\
\mathbb M\Big\{\prod_{j=1}^{2n}E(s_{\si(j)},p_{\si(j)})\Big\}.
$$
Here, the summation extends over all possible permutations
$\si:\{1,\ldots,2n\}\to\{1,\ldots,2n\}$.  The symmetry of the last
product above allows us to write
$$
c_n=\sum_{\si}\int_{\Delta_{2n}(t)} ds_ {1,2n}
\int_{\bbR^{2nd}} \bd p_{1,2n}\prod_{j=1}^n\left[\hat
R(p_{\si(2j-1)})\delta(p_{\si(2j-1)}+p_{\si(2j)})\right]
\mathbb M\Big\{\prod_{j=1}^{2n}E(s_j,p_j)\Big\}.
$$
Using the independence of
the increments of a Brownian motion, and performing the expectation we
conclude that  (with $s_{2n+1}=0$)
\begin{equation}
\label{080109}
c_n=
\sum_{\si}\int_{\Delta_{2n}(t)} ds_ {1,2n}
\int_{\bbR^{2nd}} \bd p_{1,2n}\prod_{j=1}^n\left[\hat
R(p_{\si(2j-1)})\delta(p_{\si(2j-1)}+p_{\si(2j)})\right]
S_n(s_{1,2n},p_{1,2n}),
\end{equation}
where
$$
S_n(s_{1,2n},p_{1,2n}):=\prod_{j=1}^{2n}  
\exp \left\{(s_{2n-j+1}-s_{2n-j})\left(-\frac{i}{2} \Big|\sum_{m=1}^j
    p_{m}\Big|^2+
i\xi\cdot \sum_{m=1}^j p_{m}\right) \right\}.
$$
Note that (\ref{021802abbb}) and (\ref{080109}) are very similar, 
making (\ref{010811}) ``plausible'', except for the summation taken
over all permutations $\sigma$ in (\ref{080109}), as opposed to the
summation over all pairings ${\cal F}$ in the left side of
(\ref{010811}). As we will see, the difference in the two summations
is responsible for the factor $1/(2n)!!$ in (\ref{010811}).  

To
reconcile the two summations, let $\Pi(2n)$ be the set of all
permutations of $\{1,\ldots,2n\}$, and define the mapping ${\mathfrak
  f}:\Pi(2n)\to {\mathfrak F}_{2n}$ as follows.  Given a permutation
$\si$, we let ${\mathfrak f}(\si)$ be the following pairing: a pair
$(\ell,r)$, with~$\ell<r$ is in ${\mathfrak f}(\si)$ iff there exists $j$
such that $\ell=\si(2j-1)$ and $r=\si(2j)$, or $\ell=\si(2j)$ and
$r=\si(2j-1)$. In other words, we start with the simple 
pairing~$(1,2)(3,4)\dots(2n-1,2n)$ and map it by $\sigma$ to the pairing
\begin{equation}
\label{052811}
(\sigma(1),\sigma(2))(\sigma(3),\sigma(4))\dots(\sigma(2n-1),\sigma(2n)),
\end{equation}
with a slight abuse of notation, as it is possible that
$\sigma(2j-1)>\sigma(2j)$.  Observe that if ${\cal F}={\mathfrak f}(\si)$
then
\begin{eqnarray}
\label{020811}
&&
\int_{\bbR^{2nd}} S_n(s_{1,2n},p_{1,2n}) \prod_{j=1}^n\left[\hat
R(p_{\si(2j-1)})\delta(p_{\si(2j-1)}+p_{\si(2j)})\right] \bd p_{1,2n}\\
&&
=\int_{\bbR^{nd}} S_n(s_{1,2n},p_{1,2n}) \left[\prod_{\ell\in{\cal L}({\cal F})}\hat
R(p_{\ell})\bd p_{\ell}\right] .\nonumber
\end{eqnarray}
On the other hand, given a pairing ${\cal F}\in {\mathfrak
  F}_{2n}$:
\begin{equation}
\label{030811}
 {\cal F}:=\{(\ell_1,r_1),\ldots,(\ell_n,r_n)\},
\end{equation}
with $1=\ell_1<\ell_2<\ldots<\ell_n$, we may define the corresponding
permutation ${\mathfrak g}({\cal F})\in \Pi(2n)$ 
as~$(\ell_1,r_1,\ell_2,r_2,\dots,\ell_n,r_n)$.
This defines the mapping ${\mathfrak g}:{\mathfrak F}_{2n}\to
\Pi(2n)$ such that
$$
{\mathfrak f}\left({\mathfrak g}({\cal F})\right)={\cal F},\quad \forall{\cal F}\in  {\mathfrak
  F}_{2n}.
$$
Thus,  the mapping ${\mathfrak f}$ is onto. 
Next, suppose that 
$\si={\mathfrak g}({\cal F})$ and ${\cal F}$ is given by \eqref{030811}.
Note that any permutation $\si'$ obtained from $\si$ by a transposition
of $\ell_j$ and~$r_j$, as well as by permuting in the same fashion 
$\ell_1,\ldots\ell_n$ and $r_1,\dots,r_n$, 
satisfies
$
{\mathfrak f}(\si')={\cal F}.
$
For each permutation $\si={\mathfrak g}({\cal F})$ there exist   $2^nn!=(2n)!!$ different
permutations obtained in that way. Writing ${\cal F}_\si=\mathfrak f(\sigma)$ we obtain from \eqref{080109} and
\eqref{020811} 
\begin{eqnarray*}
&&\!\!\!\!\!\!\!\frac{c_n}{(2n)!!}
=\frac{1}{(2n)!!}\int_{\Delta_{2n}(t)} ds_ {1,2n}
\int_{\bbR^{2nd}} S_n(s_{1,2n},p_{1,2n}) \prod_{j=1}^n\left[\hat
R(p_{\si(2j-1)})\delta(p_{\si(2j-1)}+p_{\si(2j)})\right]\bd p_{1,2n}\nonumber
\\
&&
=\frac{1}{(2n)!!}\sum_{\sigma} 
\int_{\Delta_{2n}(t)} ds_ {1,2n} \int \left[\prod_{\ell\in{\cal L}({\cal F}_\sigma)}S_n(s_{1,2n},p_{1,2n})\hat
R(p_{\ell})\bd p_{\ell}\right] 
\\
&&
=\sum_{\cal F}\int_{\Delta_{2n}(t)} ds_ {1,2n}\int \left[\prod_{\ell\in{\cal L}({\cal F})}S_n(s_{1,2n},p_{1,2n})\hat
R(p_{\ell})\bd p_{\ell}\right].
\end{eqnarray*}
Comparing to (\ref{021802abbb}), we conclude
that \eqref{010811} holds, finishing the proof of (\ref{010811})
and thus also of Proposition~\ref{prop010811}.
$\Box$

\section{The finite effective potential regime}
\label{sec3}

In this section, we present the proofs of Theorems~\ref{thm-may2302}
and~\ref{main1}, both of which hold when the effective potential $R_*<+\infty$,
that is, $\gamma>3-d$, and the non-trivial behavior takes 
place at times of the order $t\sim\eps^{-2}$.

Let us add the weak coupling limit to the representation in 
Proposition~\ref{prop010811}.  
Recall that, the wave function $\psi^{(\eps)}$ is the solution of
\begin{eqnarray}\label{schr1-eps}
&&i\pdr{\psi^{(\eps)}}{t}+\frac{1}{2}\Delta\psi^{(\eps)}-
\eps V(x)\psi^{(\eps)}=0,\\
&&\psi^{(\eps)}(0,x)=\eps^{d\beta/2}\psi_0(\eps^\beta x),\nonumber
\end{eqnarray}
with $\beta\ge 0$, and the
compensated wave function is given by \eqref{def-zeta-eps1}. 
We may now apply Proposition~\ref{prop010811}, replacing the random potential
$V\to\eps V$, and the time $t\to t/\eps^2$ in (\ref{c-n}) and~(\ref{c-n1}). 
Using, in addition, the
representation (\ref{schoen1abis}) 
for the power spectrum $\hat R(p)$ turns~(\ref{aug2710}), when $\beta=0$,
into 
\begin{equation}
\label{012411}
\bar \zeta_\eps(t,\xi):=\bbE \hat\zeta_\eps(t,\xi)=\hat \psi_0(\xi)
\sum_{n=0}^{+\infty}\frac{(-1)^nc_{n,\eps}(t,\xi)}{(2n)!!},\quad \forall\, (t,\xi)\in\bbR^{1+d},
\end{equation}
where
\begin{equation}
\label{nov1902}
c_{n,\eps}(t,\xi):=\eps^{2n}\mathbb
M\Big\{\int_1^{+\infty}\farc{s(\la) d\lambda}{\lambda^\gamma}\int_0^{t/\eps^2}ds
\int_0^{t/\eps^2} ds'\int_{\bbR^d}E(s,i,p,\xi) E(s',-i,p,\xi)
e^{-\la^2|p|^2/2}\bd p\Big\}^n
\end{equation}
and $E(s,i,p,\xi)$ is given by \eqref{V}. On the other hand, when $\beta>0$,
we have
\begin{equation}
\label{012411c1bis}
\bar \zeta_\eps(t,\xi)=\hat
\psi_0(\xi)\sum_{n=0}^{+\infty}\frac{(-1)^n}{(2n)!!}c_{n,\eps}(t,\eps^{\beta}
\xi).
\end{equation}
 
\subsection*{A formal analysis for $\xi=0$}

Before we embark on the proof of the main results,
we note that, say, when $\beta=0$, we may
use the definition (\ref{V}) of $E(s,i,p,\xi)$ to re-sum 
the series (\ref{012411}). This gives
\begin{eqnarray}
\label{c-n2a}
&&
\bar\zeta_\eps(t,\xi)
=\hat\psi_0(\xi)\mathbb M\left\{\exp\left\{-\frac{\eps^2}{2}
    \int_1^{+\infty}\farc{s(\la)
      d\lambda}{\lambda^\gamma}\int_0^{t/\eps^2}ds\int_0^{t/\eps^2}ds'\right.\right.\\
&&~~~~~~~~~~
\left.\left.\times\vphantom{e^{\int_0^t}} \int_{\bbR^d}e^{-\la^2|p|^2/2}\exp\left\{i\left(B^{(\xi)}_s-B^{(\xi)}_{s'}\right)\cdot p\right\}\right\}\bd p \right\},
\nonumber
\end{eqnarray}
with
\begin{equation}
\label{B-x}
B^{(\xi)}_s:=\sqrt{i}B_s+\xi s.
\end{equation}
Performing the integration over the $p$ variable we obtain the following representation of the averaged
compensated wave function when $\beta=0$ (the microscopic initial condition). 
\begin{prop}
\label{012502}
Suppose that $\psi_0\in {\cal S}(\bbR^d)$, then
\begin{eqnarray}
\label{c-n2}
&&
\bar\zeta_\eps(t,\xi)
=\hat\psi_0(\xi)\mathbb
M\exp\left\{-\frac{1}{(2\pi)^{d/2}}\int_1^{+\infty}Z_{\eps}(t,\la,\xi)\frac{s(\la)d\la}{\la^{\ga+d}}
  \right\}.
\end{eqnarray}
Here
\begin{equation}
\label{Z-e}
Z_{\eps}(t,\la,\xi):=\frac{\eps^2}{2}  \int_0^{t/\eps^2}ds\int_0^{t/\eps^2}ds'
\exp\left\{-\frac{1}{2\la^2}\left(B^{(\xi)}_s-B^{(\xi)}_{s'}\right)^2\right\}
\end{equation}
and 
 $a^2:=\sum_{j=1}^d a_j^2$ for any $a=(a_1,\ldots,a_d)\in\mathbb C^d$.
\end{prop}
It is straightforward to write down the analog for $\beta>0$ by replacing 
$\xi$ by $\eps^\beta\xi$ in (\ref{012502}) and~(\ref{Z-e}).

Representation \eqref{c-n2} is quite simple and elegant, and seems a natural starting point
for the proof of the main results of this paper. However, because of the
double exponential appearing in (\ref{c-n2}) and (\ref{Z-e}), we were unable to
find a simple way  
to pass to the limit~$\eps\to 0+$ in~(\ref{c-n2}). As we show below, one can relatively easily pass to the limit
in (\ref{Z-e}), at least when $\xi=0$, and obtain a   limit for $Z_\eps(t,\la,0)$, but even for $\xi=0$
the justification of 
interchanging the limit and the expectation in (\ref{c-n2}) eluded us. This strategy for $\xi\neq 0$ seems
not to be simpler than what is done in the rest of the paper, and the justification of interchanging the limit and the expectation
seems non-trivial, so 
we do not pursue this route here.

We now
describe how one passes to the limit in (\ref{Z-e}) 
in the simplest case $\xi=0$, as this gives a relatively quick formal way to the general answer.  
%Representation \eqref{c-n2} leads to an
%informal argument indicating that Theorem \ref{thm-may2302}
%holds for $\xi=0$, for all times $t\ge0$. 
%Denote for brevity sake $
We will show that~$Z_{\eps}(t,\la):=Z_{\eps}(t,\la,0)$ has
a deterministic limit as $\eps\to 0$: 
\begin{equation}\label{aug2402}
Z_\eps(t,\la)\to \bar Z(t,\la):=-\farc{2i\lambda^2t}{d-2}.
\end{equation}
Recall that our analysis holds in dimension $d>2$.
This is in agreement with (\ref{232611}), as
\begin{equation}
\label{may2332bis2}
r(0)=
{\frac{2}{(2\pi)^{d/2}(d-2)}}\int_1^\infty
\frac{ s(\lambda)  d\la}
{\la^{\ga+d-2}}.
\end{equation}
To show (\ref{aug2402}), first note that
\begin{eqnarray}
\label{011506}
&&\bbM Z_{\eps}(t,\la)=\eps^2 \int_0^{t/\eps^2}ds\int_0^{s}
\bbM\exp\left\{-\frac{i|B_{s'}|^2}{2\la^2}\right\}ds'
\\ &&~~~~~~~~~~~~~
=\frac{\eps^2}{(2\pi)^{d/2}} \int_0^{t/\eps^2}ds\int_0^{s}ds'
\int_{\bbR^d}\exp\left\{-\frac{|y|^2}{2}
\left(1+\frac{is'}{\la^2}\right)\right\}dy\nonumber\\ 
&&~~~~~~~~~~~~~=\eps^2\lambda^2\int_0^{t/\eps^2}ds\int_0^{s/\lambda^2}
\farc{ds'}{(1+is')^{d/2}}=\eps^2\lambda^2
\int_0^{t/(\la^2\eps^2)}\Big(\frac{t}{\eps^2}-\la^2 s'\Big) 
\farc{ds'}{(1+is')^{d/2}}\nonumber\\
&&~~~~~~~~~~~~~\to \lambda^2t\int_0^\infty \farc{ds'}{(1+is')^{d/2}}=-\farc{2i\lambda^2t}{d-2},
\nonumber
\end{eqnarray}
so that 
\begin{equation}\label{aug2404}
\bbM Z_\eps(t,\la)\to \bar Z(t,\la),
\end{equation}
given by (\ref{aug2402}).
%Using an elementary formula
%\begin{equation}
%\label{031506}
%%\int_{\bbR}\exp\left\{-(\beta x)^2-iqx\right\}dx
%%=\frac{\sqrt{\pi}}{\beta}\exp\left\{-\frac{q^2}{4\beta^2}\right\}
%\int_{\bbR}\exp\left\{-(\beta x)^2\right\}dx
%=\frac{\sqrt{\pi}}{\beta}%\exp\left\{-\frac{q^2}{4\beta^2}\right\}
%\end{equation}
%that holds for any $\beta\in\mathbb C$ such that ${\rm
%  Re}\,\beta>0$ and ${\rm
%  Re}\,\beta^2>0$ we conclude that the utmost right hand side of
%\eqref{011506} equals 
%\begin{equation}
%\label{011506ab}
%\lim_{\eps\to0+}(\eps\la)^2\int_0^{t/\eps^2}ds\int_0^{s}\frac{ds'}{(1+is')^{d/2}}= t\la^2\kappa(0,\la),
%\end{equation}
%where $\kappa(0,\la)$ is given by \eqref{kappa}. 

Next, we look at the second moment
\begin{equation}
\label{011506a}
\bbM| Z_{\eps}(t,\la)|^2=\frac{\eps^4}{4}
\int_0^{t/\eps^2}\!\!\!\!\ldots \int_0^{t/\eps^2}
\bbM\exp\left\{-\frac{i}{2\la^2}\left[\left|B_{s_1}-B_{s_2}\right|^2-\left|B_{s_3}-B_{s_4}\right|^2\right]\right\}ds_{1,4}.
\end{equation}
In the ensuing calculation we show that
\begin{equation}
\label{011506aa}
\lim_{\eps\to0+}\bbM| Z_{\eps}(t,\la)|^2=\bar Z(t,\la)^2,
%\left(\vphantom{\int_0^1}\la^2 t|\kappa(0,\la)|\right)^2
%=\left|\lim_{\eps\to0+}\bbM Z_{\eps}(t,\la)\right|^2,
\end{equation}
which, combined with \eqref{aug2404},
proves that
\begin{equation}\label{aug2408}
\lim_{\eps\to0+}Z_{\eps}(t,\la)=\bar Z(t,\la) %t\la^2\kappa(0,\la)
\end{equation}
in the $L^2$ sense, with respect to the randomness. 
As we have mentioned, this argument does not allow us to
exchange the limit, as $\eps\to0+$, and the expectation in \eqref{c-n2}.
If we formally do this, using (\ref{aug2408}), we conclude that~\eqref{232611} holds for $\xi=0$. Making this argument fully rigorous, including
for $\xi\neq 0$,
is essentially what is done in the rest of this paper

To show that \eqref{011506aa} holds, we follow the
argument made in Section \ref{sec2}. We may use
\eqref{011506a} to write
\[
\bbM| Z_{\eps}(t,\la)|^2=I_{1,\eps}+I_{2,\eps}+I_{3,\eps},
\]
with
\begin{eqnarray}
\label{021506}
&&
I_{1,\eps}:=2\eps^4
\int_{\Delta_4(t/\eps^2)}
\bbM\exp\left\{-\frac{i}{2\la^2}\left[\left|B_{s_1}-B_{s_2}\right|^2
-\left|B_{s_3}-B_{s_4}\right|^2\right]\right\}ds_{1,4},\nonumber\\
&&
I_{2,\eps}:=2\eps^4
\int_{\Delta_4(t/\eps^2)}
\bbM\exp\left\{-\frac{i}{2\la^2}\left[\left|B_{s_1}-B_{s_3}\right|^2
-\left|B_{s_2}-B_{s_4}\right|^2\right]\right\}ds_{1,4},\nonumber\\
&&
\\
&&
I_{3,\eps}:=2\eps^4
\int_{\Delta_4(t/\eps^2)}
\bbM\exp\left\{-\frac{i}{2\la^2}\left[\left|B_{s_1}-B_{s_4}\right|^2
-\left|B_{s_2}-B_{s_3}\right|^2\right]\right\}ds_{1,4},\nonumber
\end{eqnarray}
where $\Delta_4(t/\eps^2):=[t/\eps^2\ge s_4\ge \ldots \ge s_1\ge0]$ and $ds_{1,4}:=ds_1ds_2ds_3ds_4$.
Let 
$$
\Delta'_4(t/\eps^2):=\left[\sum_{j=1}^4\tau_j\le t/\eps^2, \,\tau_j\ge
0,\,j=1,\ldots,4\right].
$$
A
  direct calculation, as in the case of the first
  moment of $Z_{\eps}(t,\la) $, shows that
\begin{equation}
\label{041505}
I_{1,\eps}=
2(\eps\la)^4
\int_{\Delta_4'(t/\eps^2)}d\tau_{1,4}\prod_{j=1}^2\frac{1}{(1+(-1)^ji\tau_{2j})^{d/2}}\to(\bar Z(t,\la))^2,
%\left(\vphantom{\int_0^1}\la^2 t|\kappa(0,\la)|\right)^2,
\end{equation}
as $\eps\to0$.
Then,  \eqref{011506aa} follows, provided we  show that
\begin{equation}
\label{012906}
\lim_{\eps\to0+}I_{j,\eps}=0,\quad j=2,3.
\end{equation}
Note that
\begin{eqnarray}
\label{041505a}
&&I_{2,\eps}=
\frac{2\eps^4}{(2\pi)^{3d/2}}
\int_{\Delta_4'(t/\eps^2)}d\tau_{1,4} \int_{\bbR^{3d}}\exp\left\{-\frac{1}{2}((I_{3d}+iD)y,y)_{\bbR^{3d}}\right\}dy_{1,3},\nonumber
\end{eqnarray}
with $y:=[y_1,y_2,y_3]$, $I_d$  the $d\times d$ identity matrix
and $D$  the $3d\times 3d$ block matrix 
 obtained
from 
\begin{equation}\label{sep202}
D_r=\left[
\begin{array}{ccc}
\dfrac{\tau_2}{\la^2}&\dfrac{(\tau_2\tau_3)^{1/2}}{\la^2}&0\\
\dfrac{(\tau_2\tau_3)^{1/2}}{\la^2}&0&-\dfrac{(\tau_3\tau_4)^{1/2}}{\la^2}\\
0&-\dfrac{(\tau_3\tau_4)^{1/2}}{\la^2}&-\dfrac{\tau_4}{\la^2}
\end{array}
\right]
\end{equation}
by replacing each entry of $D_r$ by the respective $d\times d$ diagonal block. 
To evaluate the $y$-integral in \eqref{041505a} we use the following elementary
result.
\begin{lemma}
\label{lm022411}
Suppose that $f:\mathbb C^{N}\to \mathbb C$ is a holomorphic function
such that there exists $C>0$, for which
\begin{equation}
\label{062411}
|f(z_1,\ldots,z_N)|\le C\exp\left\{C\sum_{j=1}^N|z_j|\right\},\quad
(z_1,\ldots,z_N)\in\mathbb C^N
\end{equation}
and $A$ is a 
 symmetric $N\times N$-matrix  with
eigenvalues $\la_1\ge \la_2\ge \ldots \ge \la_N$. 
Then,
\begin{eqnarray}
\label{052411}
&&\int_{\bbR^N}\exp\left\{
-\frac{1}{2}\left((I_N+zA)x,x\right)_{\bbR^N}\right\}f(x)dx\\
&&~~~~~~~~=[{\rm
det}(I_N+zA)]^{-1/2}\int_{\bbR^N}\exp
\left\{-\frac{|x|^2}{2}\right\}f\left((I_N+zA)^{-1/2}x\right)dx,\nonumber
\end{eqnarray}
for all $z\in \mathbb C$ such that ${\rm Re}(1+z\la_j)>0$, $j=1,\ldots,N$.
\end{lemma}
The formula (\ref{052411})
obviously holds for $z$ real. It can be extended
to the set in question 
by the  analytic continuation argument.

Using the above lemma  we conclude that
 \begin{equation}\label{sep204}
I_{2,\eps}=
2\eps^4
\int_{\Delta_4'(t/\eps^2)}{\rm det}(I_3+iD_r)^{-d/2}d\tau_{1,4}.
\end{equation}
A direct calculation yields
 \begin{equation}\label{sep206}
{\rm det}(I_3+iD_r)=1+\frac{\tau_2\tau_4+\tau_3\tau_2+\tau_3\tau_4}{\la^4}+\frac{i(\tau_2-\tau_4)}{\la^2}.
\end{equation}
Hence, we have
 \begin{equation}\label{sep208}
|I_{2,\eps}|\le 
C\eps^4
\int_{\Delta_4'(t/\eps^2)}\left\{1+\frac{\tau_2\tau_4+\tau_3\tau_2+\tau_3\tau_4}{\la^4}+\frac{|\tau_2-\tau_4|}{\la^2}\right\}^{-d/2}d\tau_{1,4}.
 \end{equation}
Changing variables $\tau_j':=\eps^2\tau_j$, $j=1,3$, we  obtain
 \begin{equation}\label{sep210}
|I_{2,\eps}|\le 2^{d/2+1}\int_{\Delta_2'(t)}d\tau_1d\tau_3\int_0^{+\infty}\int_0^{+\infty}
\left\{1+\frac{\eps^{-2}\tau_3(\tau_2+\tau_4)+\tau_2\tau_4}{\la^4}+\frac{|\tau_2-\tau_4|}{\la^2}\right\}^{-d/2}d\tau_2d\tau_4\to0,
 \end{equation}
as $\eps\to0+$, since the integrand is bounded by the function
$
\left\{1+\tau_2\tau_4\la^{-4}+|\tau_2-\tau_4|\la^{-2}\right\}^{-d/2}
$
that is absolutely integrable in $[0,+\infty)^2$ for $d\ge3$.
Similarly, we have
\begin{equation}
\label{041505aa}
I_{3,\eps}=
2\eps^4
\int_{\Delta_4'(t/\eps^2)}{\rm det}(I_3+iE_r)^{-d/2}d\tau_{1,4},
\end{equation}
with 
 \begin{equation}\label{sep212}
E_r=\left[
\begin{array}{ccc}
\dfrac{\tau_2}{\la^2}&\dfrac{(\tau_2\tau_3)^{1/2}}{\la^2}&\dfrac{(\tau_2\tau_4)^{1/2}}{\la^2}\\
\dfrac{(\tau_2\tau_3)^{1/2}}{\la^2}&0&\dfrac{(\tau_3\tau_4)^{1/2}}{\la^2}\\
\dfrac{(\tau_2\tau_4)^{1/2}}{\la^2}&\dfrac{(\tau_3\tau_4)^{1/2}}{\la^2}&\dfrac{\tau_4}{\la^2}
\end{array}
\right].
\end{equation}
One can easily obtain
 \begin{equation}\label{sep214}
{\rm det}(I_3+iE_r)=1+\frac{\tau_2\tau_3+\tau_3\tau_4}{\la^4}+\frac{i(\tau_2+\tau_4)}{\la^2}.
\end{equation}
It follows that
\begin{equation}\label{sep216}
|I_{3,\eps}|\le 
2^{d/2+1}\eps^4
\int_{\Delta_4'(t/\eps^2)}\left\{1+\frac{\tau_3\tau_2+\tau_3\tau_4}{\la^4}+\frac{\tau_2+\tau_4}{\la^2}\right\}^{-d/2}d\tau_{1,4}.
 \end{equation}
Changing variables $\tau_j':=\eps^2\tau_j$, $j=1,2$ we conclude that
\begin{equation}\label{sep218}
|I_{3,\eps}|\le C\int_{\Delta_2'(t)}d\tau_1d\tau_2\int_0^{t/\eps^2}d\tau_3
\left(\int_0^{t/\eps^2}\left\{A_\eps+B\tau_4\right\}^{-d/2}d\tau_4\right),
\end{equation}
with 
\begin{equation}\label{sep220}
A_\eps:=1+(\la^2\eps)^{-2}\tau_3\tau_2+(\la\eps)^{-2}\tau_2,\quad B:=\frac{\tau_3}{\la^4}+\frac{1}{\la^2}.
\end{equation}
Substituting $\tau_4':=A_\eps+B\tau_4$ in the last integral we obtain  
\begin{eqnarray}\label{sep222}
&&|I_{3,\eps}|\le C\int_{\Delta_2'(t)}d\tau_1d\tau_2\int_0^{t/\eps^2}B^{-1}A_\eps^{1-d/2}d\tau_3
\\
&&~~~~~~\le 
{C\la^4}\log \left[\frac{t}{(\eps\la)^{2}}+1\right]
\int_{\Delta_2'(t)}\left(1+(\la\eps)^{-2}\tau_2\right)^{1-d/2}d\tau_1d\tau_2\to 0,\nonumber
\end{eqnarray}
as $\eps\to0+$, since $d>2$. Thus, \eqref{012906} follows, and the proof of (\ref{aug2408}) is complete. As we have mentioned,
unfortunately, the passage to the limit in (\ref{Z-e}) for $\xi\neq 0$ and justification of the interchange of the limit and expectation 
in (\ref{c-n2}) do not seem simpler than what is done in the rest of the paper, so we do not use (\ref{c-n2}) and (\ref{Z-e}) below.
%\\
%&&
%2^{d/2+1}\int_{\Delta_2'(t)}d\tau_1d\tau_2\to0,
% 
%

\subsection*{A uniform bound on $c_{n,\eps}(t,\xi)$} 

The main step in the proof of Theorems~\ref{thm-may2302}
and~\ref{main1} is the following uniform bound on $c_{n,\eps}$ that allows
us to pass to the limit in representations (\ref{012411}) and 
(\ref{012411c1bis}).
\begin{prop}
\label{prop1811}
Suppose that $d\ge3$ and $\gamma>3-d$.
Then, there exists $C>0$ such that
for all $n\ge0$, $\eps\in(0,1]$, $\xi\in\bbR^d$ and $t\ge0$ we have
\begin{equation}
\label{121711a}
|c_{n,\eps}(t,\xi)|\le n!(Ct)^n.
\end{equation}
\end{prop}
 
\subsection*{An alternative representation for $c_{n,\eps}(t,\xi)$} 
 
Both in the proof of  Proposition~\ref{prop1811}, and in passing to the
limit $\eps\to 0$ in $c_{n,\eps}(t,\xi)$, it will be convenient for us
to use an expression different from \eqref{nov1902}. Let us first introduce 
some notation. Given a permutation $\sigma$ of $\{1,\dots,2n\}$,
we have the corresponding pairing in ${\mathfrak F}_{2n}$ defined by 
\begin{equation}
\label{021711}
{\mathfrak f}(\si)=\{(\ell_1,r_1),\ldots,(\ell_n,r_n)\},
\end{equation}
with
\begin{equation}\label{nov2002}
(\ell_{k},r_k):=\left\{
\begin{array}{ll}
(\si(2k-1),\si(2k)),& \mbox{ if }\si(2k-1)<\si(2k),\\
&\\
(\si(2k),\si(2k-1)),& \mbox{ if }\si(2k)<\si(2k-1).
\end{array}
\right.
\end{equation}
We may then define a $2nd\times 2nd$ symmetric non-negative matrix 
$A_{\si}(\tau,\lambda)$ corresponding to
the quadratic form
\begin{equation}
\label{Phi}
\Phi_{\si}(y)=\left(A_{\si}(\tau,\lambda)y,y\right)_{\bbR^{2dn}}=
\sum_{k=1}^{n}\frac{1}{\lambda_{r_k}^2}\left|\sum_{j=\ell_k+1}^{r_k}y_{j}\tau_j^{1/2}\right|^2
\end{equation}
for $y=(y_1,\ldots,y_{2n})\in \bbR^{2dn}$ and $\tau_j:=s_j-s_{j-1}$,
$j=1,\ldots,2n$ with $s_0:=0$. 
In order to  describe  the matrix $A_\si(\tau,\la)$ more
explicitly, we introduce some terminology.
 
For each $j\in\{2,\ldots,2n\}$ define 
\begin{equation}
\label{aj}
a_{j,j}:=\sum_k\!\vphantom{1}'\frac{1}{\la_{r_k}^2},
\end{equation}
 where the summation extends over those $k$-s, for which
$
\ell_k< j\le r_k.
$
Given $m<j$ we let
\begin{equation}
\label{aj1}
a_{m,j}=a_{j,m}=\sum_k\!\vphantom{1}'\frac{1}{\la_{r_k}^2},
\end{equation}
with the summation extending over  those $k$-s, for which
$
\ell_k< m< j\le r_k.
$
We also let 
\begin{equation}
\label{aj2}
a_{1,j}=a_{j,1}=0\quad\mbox{ for all }\quad j\in\{1,\ldots,2n\}.
\end{equation}
Then, the matrix $A_\si$ has the form
\begin{equation}
\label{A}
A_{\si}(\tau,\la)=\left[
\begin{array}{lccccc}
0&0&0&0&0\ldots&0\\
0&I_da_{2,2}\tau_2&I_da_{23}\tau_2^{1/2}\tau_3^{1/2}&I_da_{2,4}\tau_2^{1/2}\tau_4^{1/2}&\ldots&I_da_{2,2n}\tau_2^{1/2}\tau_{2n}^{1/2}\\
0&I_da_{3,2}\tau_{3}^{1/2}\tau_2^{1/2}&I_da_{3,3}\tau_3&I_da_{34}\tau_3^{1/2}\tau_4^{1/2}&\ldots&I_da_{3,2n}\tau_3^{1/2}\tau_{2n}^{1/2}\\
\ldots&\ldots&\ldots&\ldots&\ldots&\ldots\\
0&I_da_{2n,2}\tau_{2n}^{1/2}\tau_2^{1/2}&I_da_{2n,3}\tau_{2n}^{1/2}\tau_3^{1/2}&I_da_{2n,4}\tau_{2n}^{1/2}\tau_4^{1/2}&\ldots&I_da_{2n,2n}\tau_{2n}
\end{array}
\right].
\end{equation}
Here,  $0$ and $I_d$ are the null and
identity $d\times d$ matrices respectively.

We will show the following.
\begin{prop}\label{lem-jun2704}
We have
\begin{eqnarray}
\label{031911}
&&
c_{n,\eps}(t,\xi)
= \left(\frac{\eps^2}{(2\pi)^{d/2}}\right)^n
\sum_{\si}\int_1^{+\infty}\frac{s(\la_{r_1})d\la_{r_1}}{\la_{r_1}^{\ga+d}}\ldots \int_1^{+\infty}\frac{s(\la_{r_n})d\la_{r_n}}{\la_{r_n}^{\ga+d}}\int_{\tilde\Delta_{2n}(t/\eps^2)}d\tau_{1,2n}\nonumber\\
&&
\\
&&
~~~~~~
~~~~\times{\rm det}( I_{2nd}+iA_{\si}(\tau,\lambda))^{-1/2}\exp
\left\{-\frac{1}{2}(C_{\si}(\tau,\la)\Xi(\tau),\Xi(\tau))_{\bbR^{2nd}}\right\},\nonumber
\end{eqnarray}
with the matrix
\begin{eqnarray}\label{nov2504}
C_\si(\tau,\la)=A_\si(\tau,\la)-{i}(I_{2nd}+iA_\si(\tau,\la))^{-1} A_{\si}^2(\tau,\la)=( I_{2nd}+iA_{\si}(\tau,\la))^{-1}A_{\si}(\tau,\la)
\end{eqnarray}
 and
\begin{equation}
\label{032411}
\Xi^T(\tau):=[\tau_1^{1/2}\xi,\ldots, \tau_{2n}^{1/2}\xi].
\end{equation}
\end{prop}
 
%\subsection*{Proof of Proposition~\ref{prop1811}}%\label{sec:proof-prop} 

\subsubsection*{The proof of Proposition~\ref{lem-jun2704}: the case $\xi=0$}

%\label{sec4.2}

We will first consider the special case $\xi=0$. For
simplicity of notation, we let $s(\lambda)=1$, as the case of a
general 
non-negative bounded
function $s(\lambda)$ is essentially identical. 
We have
\begin{equation}
\label{c-n-0}
c_{n,\eps}:=c_{n,\eps}(t,0)=\eps^{2n}\mathbb
M\left\{\int_1^{+\infty}\farc{d\lambda}{\lambda^\gamma}\int_0^{t/\eps^2}ds
\int_0^{t/\eps^2} ds'\int_{\bbR^d}
e^{i^{3/2}p\cdot(B_s-B_{s'})}
e^{-\lambda^2|p|^2/2}\bd p \right\}^n.
\end{equation}
Performing integration over $p$ in \eqref{c-n-0}, we obtain
\begin{equation}
\label{011511}
c_{n,\eps}=\eps^{2n}\mathbb M\left\{\frac{1}{(2\pi)^{d/2}}\int_1^{+\infty}\farc{d\lambda}{\lambda^{\gamma+d}}
\int_0^{t/\eps^2}ds\int_0^{t/\eps^2} ds' \exp
\left\{-\frac{i}{2\lambda^2}|B_s-B_{s'}|^2\right\} \right\}^n.
\end{equation}
This may be re-written as
\begin{eqnarray*}
&&c_{n,\eps}=\Big(\frac{\eps^{2}}{(2\pi)^{d/2}}\Big)^n 
 \int_1^{+\infty}\farc{d\lambda_1}{\lambda_1^{(\gamma+d)/2}}\dots
\int_1^{+\infty}\farc{d\lambda_{2n}}{\lambda_{2n}^{(\gamma+d)/2}}\prod_{j=1}^n\delta(\lambda_{2j-1}-\lambda_{2j})\\
&&~~~~~\times
\int_0^{t/\eps^2}ds_1\dots\int_0^{t/\eps^2} ds_{2n} \Mm\exp
\left\{-\sum_{k=1}^n\frac{i}{2\lambda_{2k}^2}|B_{s_{2k-1}}-B_{s_{2k}}|^2\right\}.
\end{eqnarray*}  
Once again, for each $(s_1,\dots,s_{2n})$ we re-arrange the times in
the increasing order and obtain
\begin{eqnarray}\label{nov1904}
&&
c_{n,\eps}=\left[\frac{\eps^2}{(2\pi)^{d/2}}\right]^n \sum_{\sigma}
\int_1^{+\infty}\farc{d\lambda_1}{\lambda_1^{(\gamma+d)/2}}\dots
\int_1^{+\infty}\farc{d\lambda_{2n}}{\lambda_{2n}^{(\gamma+d)/2}}\prod_{j=1}^n\delta(\lambda_{\sigma(2j-1)}-
\lambda_{\sigma(2j)})\\
&&
~~~~~~~~~~~~~~~~~~~~~~~\times \int_{\Delta_{2n}(t/\eps^2)}ds_{1,2n}\mathbb
M\left\{\exp
\left\{- \sum_{k=1}^{n}\frac{i}{2\lambda_{\sigma(2k)}^2}|B_{s_{\si(2k-1)}}-B_{s_{\si(2k)}}|^2\right\}
\right\}.\nonumber
\end{eqnarray}
Here, the summation extends over all possible permutations $\si$ of
the set $\{1,\ldots,2n\}$ and, as we recall: 
\[
\Delta_{2n}(t/\eps^2)=\{0\le s_{1}\le s_{2}\le\dots\le s_{2n}\le
t/\eps^2\}.
\] 
In order to evaluate the expectation with respect to the (increments)
of the Brownian motion in (\ref{nov1904}), we will use Lemma \ref{lm022411} with $f\equiv 1$ and obtain
%\begin{lemma}
%\label{lm011511}
\begin{equation}
\label{021511}
\left(\frac{1}{2\pi}\right)^{n/2}
\int_{\bbR^{n}}\exp\left\{-\frac{1}{2}(|y|^2+z(Ay,y)_{\bbR^n})\right\}dy=\left[{\rm
    det}
(I_n+zA)\right]^{-1/2}.
\end{equation}
for any $A$  an $n\times n$ symmetric, positive definite matrix and ${\rm Re}\,
z\ge0 $.

We now rewrite (\ref{nov1904}), using the independence of the
increments of the Brownian motion, 
in the form
\begin{eqnarray*}
%\label{031511}
&&\!\!\!\!\!\!\!\!
c_{n,\eps}=\left(\frac{\eps^2}{(2\pi)^{d/2}}\right)^n 
\sum_{\si}\int_1^{+\infty}\farc{d\lambda_1}{\lambda_1^{(\gamma+d)/2}}\dots
\int_1^{+\infty}\farc{d\lambda_{2n}}{\lambda_{2n}^{(\gamma+d)/2}}\prod_{j=1}^n\delta(\lambda_{\ell_j}-
\lambda_{r_j})
\int_{\tilde\Delta_{2n}(t/\eps^2)}d\tau_{1,2n}\nonumber\\
&& ~~~~~~~~~~~~~~~\times\int_{\bbR^{2nd}}\exp
\Big\{- 
\sum_{k=1}^{n}\frac{i}{2\lambda_{r_k}^2}\Big|\sum_{j=\ell_k+1}^{r_k}y_{j}\tau_j^{1/2}\Big|^2\Big\},
%\nonumber\\
%&&
%\times
\prod_{m=1}^{2n}\left[\frac{1}{(2\pi)^{d/2}}\exp
\left\{-\frac{|y_m|^2}{2}\right\}\right]dy_{1,2n}.
%=\farc{\eps^{2n}}{(2\pi)^{nd/2}}
%\sum_{\si}\int_1^{+\infty}\farc{d\lambda_1}{\lambda_1^{(\gamma+d)/2}}\dots
%\int_1^{+\infty}\farc{d\lambda_{2n}}{\lambda_{2n}^{(\gamma+d)/2}}\prod_{j=1}^n
%\delta(\lambda_{\ell_j}-\lambda_{r_j})
%\int_{\tilde{\Delta}_{2n}(t/\eps^2)}d\tau_{1,2n}\\
%&&~~~~~~~~~~~~~~~~\times
% \int_{\bbR^{2nd}}\exp
%\Big\{- \sum_{k=1}^{n}\frac{i}{2\lambda_{r_k}^2}\Big|\sum_{j=\ell_k+1}^{r_k}y_{j}\Big|^2\Big\}
%\prod_{m=1}^{2n}\Big[\frac{1}{(2\pi\tau_{m})^{d/2}}\exp
%\Big\{-\frac{|y_m|^2}{2\tau_m}\Big\}\Big]dy_{1,2n},\nonumber
\end{eqnarray*}
with 
$$
\tilde{\Delta}_n(t):=\left[(\tau_1,\ldots,\tau_n):\,\sum_{j=1}^n\tau_j\le t,\,\tau_j\ge0,\,j=1,\ldots,n\right],
$$
We set here $d\tau_{1,2n}:=d\tau_1\ldots d\tau_{2n}$ and $dy_{1,2n}=dy_1\ldots
dy_{2n}$, $s_0:=0$.
%Changing the $y_j$-variables to $dy_j':=\tau_j^{-1/2}dy_j $, we
%get
%\begin{eqnarray*}
%%\label{041511}
%&&
%c_{n,\eps}
%=\left(\frac{\eps^2}{(2\pi)^{d/2}}\right)^n 
%\sum_{\si}\int_1^{+\infty}\farc{d\lambda_1}{\lambda_1^{(\gamma+d)/2}}\dots
%\int_1^{+\infty}\farc{d\lambda_{2n}}{\lambda_{2n}^{(\gamma+d)/2}}\prod_{j=1}^n\delta(\lambda_{\ell_j}-
%\lambda_{r_j})
%\int_{\tilde\Delta_{2n}(t/\eps^2)}d\tau_{1,2n}\nonumber\\
%&& ~~~~~~~~~~~~~~~\times\int_{\bbR^{2nd}}\exp
%\Big\{- 
%\sum_{k=1}^{n}\frac{i}{2\lambda_{r_k}^2}\Big|\sum_{j=\ell_k+1}^{r_k}y_{j}\tau_j^{1/2}\Big|^2\Big\}
%%\nonumber\\
%%&&
%%\times
%\prod_{m=1}^{2n}\left[\frac{1}{(2\pi)^{d/2}}\exp
%\left\{-\frac{|y_m|^2}{2}\right\}\right]dy_{1,2n}.
%\end{eqnarray*}
Using Lemma \ref{lm022411}, we obtain
\begin{eqnarray}
\label{051511}
&&c_{n,\eps}
=\left(\frac{\eps^2}{(2\pi)^{d/2}}\right)^n 
\sum_{\si}\int_1^{+\infty}\farc{d\lambda_1}{\lambda_1^{(\gamma+d)/2}}\dots
\int_1^{+\infty}\farc{d\lambda_{2n}}{\lambda_{2n}^{(\gamma+d)/2}}\prod_{j=1}^n
\delta(\lambda_{\ell_j}-\lambda_{r_j})\\
&&~~~~~~~~~~~~~~~~~~~~~~~~
\times\int_{\tilde\Delta_{2n}(t/\eps^2)}\left[{\rm det}(I_{2nd}+iA_{\si}(\tau,\lambda))\right]^{-1/2}d\tau_{1,2n}.\nonumber
\end{eqnarray}
This is exactly (\ref{031911}) for $\xi=0$.
%Here,
%$A_{\si}(\tau,\lambda)$ is definedthe $2nd\times 2nd$ symmetric non-negative matrix corresponding to
%the quadratic form
%\begin{equation}
%\label{Phi}
%\Phi_{\si}(y)=\left(A_{\si}(\tau,\lambda)y,y\right)_{\bbR^{2dn}}=
%\sum_{k=1}^{n}\frac{1}{\lambda_{r_k}^2}\left|\sum_{j=\ell_k+1}^{r_k}y_{j}\tau_j^{1/2}\right|^2
%\end{equation}
%for $y=(y_1,\ldots,y_{2n})\in \bbR^{2dn}$ and $\tau_j:=s_j-s_{j-1}$,
%$j=1,\ldots,2n$ with $s_0:=0$. 
 
\subsubsection*{The case of a general $\xi$}

We now extend  representation (\ref{051511})
to 
non-vanishing $\xi$. The computation is slightly more tedious 
but quite straightforward. To abbreviate, we again write $c_{n,\eps}=c_{n,\eps}(t,\xi)$.  We 
still assume for 
simplicity of the notation that $s(\lambda)\equiv 1$:
\begin{equation}
\label{c-n-1}
c_{n,\eps}=\eps^{2n}\mathbb
M\left\{\int_1^{+\infty}\farc{d\lambda}{\lambda^\gamma}\int_0^{t/\eps^2}ds
\int_0^{t/\eps^2} ds'\int_{\bbR^d}
e^{i(B^{(\xi)}_s-B^{(\xi)}_{s'})\cdot p}
e^{-\lambda^2|p|^2/2}\bd p \right\}^n.
\end{equation}
Performing integration over $p$ in \eqref{c-n-1}, we obtain
\begin{eqnarray}
\label{011911}
&&
\!\!\!\!
c_{n,\eps}=\left(\frac{\eps^{2}}{(2\pi )^{d/2}}\right)^n\mathbb
   M\Big[
\int_1^{+\infty}\frac{d\la}{\la^{\ga+d}}\int_0^{t/\eps^2}\!\!ds\int_0^{t/\eps^2}
   \!\! \!ds' \exp
\Big\{-\frac{1}{2\la^2}\big(B^{(\xi)}_s-B^{(\xi)}_{s'}\big)^2\Big\} \Big]^n\nonumber\\
&&
=\left(\frac{\eps^2}{(2\pi)^{d/2}}\right)^n 
\int_1^{+\infty}\frac{d\la_1}{\la_1^{(\ga+d)/2}}\ldots \int_1^{+\infty}\frac{d\la_{2n}}{\la_{2n}^{(\ga+d)/2}}\int_0^{t/\eps^2}ds_1\ldots \int_0^{t/\eps^2} ds_{2n}\prod_{k=1}^n\delta(\la_{2k}-\la_{2k-1}) \nonumber
\\
&&
\times\exp
\Big\{-\frac{|\xi|^2}{2}\sum_{k=1}^{n}\Big(\frac{s_{2k-1}-s_{2k}}{\la_{2k}}\Big)^{2}\Big\}
\\
&&
 \times\mathbb M\Big\{\exp
\Big\{-\frac{i}{2}\sum_{k=1}^{n}\la_{2k}^{-2}|B_{s_{2k}}-B_{s_{2k-1}}|^2-\sqrt{i}\xi\cdot
\Big[\sum_{k=1}^{n}\la_{2k}^{-2}(s_{2k}-s_{2k-1})(B_{s_{2k}}-B_{s_{2k-1}})\Big]\Big\}
\Big\}.\nonumber
\end{eqnarray}
Re-arranging again the times $s_j$ in the increasing order, we obtain
\begin{eqnarray}
\label{011911bis}
&&
\!\!c_{n,\eps}=\left(\frac{\eps^2}{(2\pi)^{d/2}}\right)^n 
\sum_{\si}\int_1^{+\infty}\frac{d\la_1}{\la_1^{(\ga+d)/2}}\ldots 
\int_1^{+\infty}\frac{d\la_{2n}}{\la_{2n}^{(\ga+d)/2}}\prod_{k=1}^n\int_{\Delta_{2n}(t/\eps^2)}d\tau_{1,2n}\delta(\la_{\si(2k)}-\la_{\si(2k-1)})
\nonumber\\
&&
\times  \exp
\left\{-\frac{|\xi|^2}{2}\sum_{k=1}^{n}\left(\frac{s_{\si(2k)}-s_{\si(2k-1)}}{\la_{\si(2k)}}\right)^{2}\right\}\mathbb M\left\{\exp
\left\{-\frac{i}{2}
\sum_{k=1}^{n}\la_{\si(2k)}^{-2}|B_{s_{\si(2k)}}-B_{s_{\si(2k-1)}}|^2\right.\right.\nonumber\\
&&
\left.\left.-\sqrt{i}
\xi\cdot\left[\sum_{k=1}^{n}\la_{\si(2k)}^{-2}(s_{\si(2k)}-s_{\si(2k-1)})(B_{s_{\si(2k)}}-B_{s_{\si(2k-1)}})\right]\right\}
\right\}.\nonumber
\end{eqnarray}
Using the formula for the joint probability  density of the random
vector $(B_{s_{\si(1)}},\ldots, B_{s_{\si(2n)}})$ leads to
\begin{eqnarray}
\label{021911}
&&
c_{n,\eps}
=\left(\frac{\eps^2}{(2\pi)^{d/2}}\right)^n \sum_{\si}\int_1^{+\infty}
\frac{d\la_{r_1}}{\la_{r_1}^{\ga+d}}\ldots \int_1^{+\infty}
\frac{d\la_{r_n}}{\la_{r_n}^{\ga+d}}\int_{\Delta_{2n}(t/\eps^2)}ds_{1,2n}\\
&&
\times\exp
\left\{-\frac{|\xi|^2}{2}\sum_{k=1}^{n}\left(\frac{s_{\si(2k)}-s_{\si(2k-1)}}{\la_{\si(2k)}}\right)^{2}\right\}
\int_{\bbR^{2dn}}\exp
\left\{-\frac{i}{2}\sum_{k=1}^{n}\la_{r_k}^{-2}\left|\sum_{j=\ell_k+1}^{r_k}y_{j}(s_j-s_{j-1})^{1/2}\right|^2\right.\nonumber\\
&&
\left.-\sqrt{i}\xi\cdot\left[\sum_{k=1}^{n}\la_{r_k}^{-2}(s_{r_k}-s_{\ell_k})\left(\sum_{j=\ell_k+1}^{r_k}y_{j}(s_j-s_{j-1})^{1/2}\right)\right]\right\}
\prod_{\ell=1}^{2n}\left[\frac{1}{(2\pi)^{d/2}}\exp
\left\{-\frac{|y_\ell|^2}{2}\right\}\right]dy_{1,2n}.\nonumber
\end{eqnarray}
Changing variables $\tau_j:=s_j-s_{j-1}$ we obtain
\begin{eqnarray}
\label{021911a}
&&
c_{n,\eps}=\left(\frac{\eps^2}{(2\pi)^{d/2}}\right)^n \sum_{\si}
\int_1^{+\infty}\frac{d\la_{r_1}}{\la_{r_1}^{\ga+d}}\ldots \int_1^{+\infty}
\frac{d\la_{r_n}}{\la_{r_n}^{\ga+d}}\int_{\tilde\Delta_{2n}(t/\eps^2)}d\tau_{1,2n}\\
&&
\times\exp
\left\{-\frac{1}{2}(A_{\si}(\tau,\lambda)\Xi(\tau)\cdot \Xi(\tau)) \right\}
\int_{\bbR^{2dn}}\exp
\left\{-\frac{1}{2}\big((I_{2nd}+iA_{\si}(\tau,\lambda))y\cdot y\big) \right\}\nonumber\\
&&
\times\exp
\left\{-\sqrt{i}(A_{\si}(\tau,\lambda)\Xi(\tau)\cdot y)  \right\}
\frac{dy_{1,2n}}{(2\pi)^{nd}}.\nonumber
\end{eqnarray}
Here,
$A_{\si}(\tau,\la)$ are the $2n d\times 2nd $  block matrices
as in \eqref{nov2010},   and $\Xi(\tau)$ is as in (\ref{032411}).
%\begin{equation}
%\label{032411bis}
%\Xi^T(\tau):=[\tau_1^{1/2}\xi,\ldots, \tau_{2n}^{1/2}\xi].
%\end{equation}
Let us re-write the $y$-integral 
using formula \eqref{052411}:
\begin{eqnarray}\label{nov2502}
&&\!\!\!\!\!\!\!\!\int_{\bbR^{2dn}}\exp
\big\{-\frac{1}{2}((I_{2nd}+i A_{\si}(\tau,\lambda))y\cdot y) \big\}
\exp
\big\{-\sqrt{i}(A_{\si}(\tau,\lambda)\Xi(\tau)\cdot y)  \big\}
\farc{dy_{1,2n}}{(2\pi)^{nd}}\nonumber\\
&&\!\!\!\!\!\!\!\!=[\hbox{det}(I_{2nd}+iA_{\si}(\tau,\lambda))]^{-1/2}\!\int\limits_{\bbR^{2nd}}
e^{-|x|^2/2}
\exp\big\{-\sqrt{i}\Big(A_{\si}(\tau,\lambda)\Xi(\tau)\cdot (I_{2nd}+iA_\si(\tau,\lambda))^{-1/2}x\Big)  \big\}
dx\nonumber\\
&&\!\!\!\!\!\!\!\!\!=
 [\hbox{det}(I_{2nd}+iA_{\si}(\tau,\lambda))]^{-1/2}
\\
&&\!\!\!\!\!\!\!\!\!
\times
\exp\big\{\farc{i}{2}
\Big((I_{2nd}+iA_\si(\tau,\lambda))^{-1/2} A_{\si}(\tau,\lambda)\Xi(\tau)\cdot
(I_{2nd}+iA_\si(\tau,\lambda))^{-1/2}A_{\si}(\tau,\lambda)\Xi(\tau)\Big) 
\big\}.
\nonumber
\end{eqnarray}
Using this in (\ref{021911a}) gives (\ref{031911}), finishing the
proof of Proposition~\ref{lem-jun2704}.~$\Box$

\subsection*{The proof of Proposition~\ref{prop1811}}

%This gives
%\begin{eqnarray}
%\label{031911bis}
%&&
%c_{n,\eps}
%= \left(\frac{\eps^2}{(2\pi)^{d/2}}\right)^n
%\sum_{\si}\int_1^{+\infty}\frac{d\la_{r_1}}{\la_{r_1}^{\ga+d}}\ldots \int_1^{+\infty}\frac{d\la_{r_n}}{\la_{r_n}^{\ga+d}}\int_{\tilde\Delta_{2n}(t/\eps^2)}d\tau_{1,2n}\\
%&&
%\times{\rm det}( I+iA_{\si}(\tau,\lambda))^{-1/2}\exp
%\left\{-\frac{1}{2}(C_{\si}(\tau,\la)\Xi(\tau),\Xi(\tau))_{\bbR^{2nd}}\right\},\nonumber
%\end{eqnarray}
%with the matrix
%\begin{eqnarray}\label{nov2504bis}
%C_\si(\tau,\la)=A_\si(\tau,\la)-{i}(I+iA_\si(\tau,\la))^{-1} A_{\si}^2(\tau,\la)=( I+iA_{\si}(\tau,\la))^{-1}A_{\si}(\tau,\la).
%\end{eqnarray}
% and
% $$
% C_{\si}(\tau,z):=( I-i z^2A_{\si}(s))^{-1}A_{\si}(s).
% $$
% Note that when $z=i$, we get
% \[
% C_{\si}(\tau,\la,i)=(I+iA_\si(\tau,\la))^{-1}A_\si(\tau,\la).
% \]
% thus the   matrix $C_\si(\tau,i)$ is non-negative and satisfies
% $C_\si(\tau,i)\le I$. 

We now use representation (\ref{031911}) for $c_{n,\eps}$
in order to obtain the bound (\ref{121711a}) in Proposition~\ref{prop1811}.
Once again, we first consider the simpler case $\xi=0$, and then the general 
case.

\subsubsection*{The estimate for $\xi=0$}

The main step in the proof is the following lower bound. 
\begin{proposition}
\label{prop011711bis}
For any permutation $\si$ we have
\begin{equation}
\label{011711bis}
|{\rm det}(I+iA_{\si}(\tau,\lambda))|
\ge \frac{1}{2^{dn/2}}\prod_{k=1}^n\left(1+ \frac{\tau_{r_k}}{\lambda_{r_k}^2}\right)^d.
\end{equation}
%where $\tau_j=s_j-s_{j-1}$. 
\end{proposition}
Before proving this proposition, let us show how it implies the
required estimate on $c_{n,\eps}$ for~$\xi=0$.
Combining (\ref{031911}) and (\ref{011711bis}) implies that there exists $C>0$ such that
\begin{equation}
\label{051511cbis}
|c_{n,\eps}|
\le C^n\eps^{2n} 
\sum_{\si}\int\limits_1^{+\infty}\frac{d\la_{r_1}}{\la_{r_1}^{\ga+d}}\ldots 
\int\limits_1^{+\infty}\frac{d\la_{r_n}}{\la_{r_n}^{\ga+d}}
\int_{\sum_{k=1}^n\tau_{\ell_k}\le t/\eps^2,
\tau_{\ell_k}\ge0}d\tau_{\ell_1}\ldots d\tau_{\ell_n}
%\nonumber\\
% &&
%\times 
\prod_{k=1}^n\int\limits_0^{+\infty}
\frac{d\tau_{r_k}}{(1+\la_{r_k}^{-2}\tau_{r_k})^{d/2}}
\end{equation}
for all $\eps>0$ and $n\ge0$.
Changing variables $\tau_{r_k}':=\la_{r_k}^{-2}\tau_{r_k}$ we obtain
\begin{eqnarray}
\label{051511cbis2}
&&
|c_{n,\eps}|
\le C^n\eps^{2n} 
\sum_{\si}\int_1^{+\infty}\frac{d\la_{r_1}}{\la_{r_1}^{\ga+d-2}}\ldots \int_1^{+\infty}\frac{d\la_{r_n}}{\la_{r_n}^{\ga+d-2}}\int_{\sum_{k=1}^n\tau_{\ell_k}\le t/\eps^2,
  \tau_{\ell_k}\ge0}d\tau_{\ell_1}\ldots d\tau_{\ell_n}\nonumber\\
  &&
  \times
  \prod_{k=1}^n\int_0^{+\infty}\frac{d\tau_{r_k}}{(1+\tau_{r_k})^{d/2}}\le
     (2n)!\farc{(Ct)^n}{n!}\le
n!(Ct)^n,
\end{eqnarray}
provided that $\ga>3-d$, so that (\ref{121711a}) holds. This
proves Proposition~\ref{prop1811} when~$\xi=0$, except for the
proof of Proposition~\ref{prop011711bis}. 

\bigskip

\subsubsection*{The proof of Proposition~\ref{prop011711bis}}

% Note that all the eigenvalues of $ I+iA_{\si}(s)$ are of the form
% $1+i\mu$, where $\mu\ge0$. Indeed suppose that $e\in\bbR^{2nd}$ is an
% eigenvector of the matrix that corresponds to the eigenvalue
% $\la$. Then
% $$
%  (I+iA_{\si}(s))y=\la y
% $$
% and, as a result,
% $$
% A_{\si}(s)y=-i(\la-1) y
% $$
% so $-i(\la-1)$ has to be a  non-negative real.

In order to describe the matrix $A_\si(\tau,\lambda)$ more explicitly, we
make a change of variables:  
\[
z_{l_k}=y_{l_k}\in\Rm^d,
\hbox{ and
}z_{r_k}=\farc{1}{\lambda_{r_k}}\sum_{j=\ell_k+1}^{r_k}y_{j}(s_j-s_{j-1})^{1/2}\in\Rm^d,
~~k=1,\dots,n,
\]
so that
\begin{equation}
\label{030412}
\left(A_{\si}(\tau,\lambda)y,y\right)_{\bbR^{2dn}}=\sum_{k=1}^n|z_{r_k}(\tau,\lambda)|^2=
(P_\sigma z,z)=(P_\sigma L(\tau,\lambda)y,L(\tau,\lambda)y).
\end{equation}
Here $P_\sigma$ is the projection matrix onto the $r_k$-components,
and $L(\tau,\lambda)$ is the matrix relating $z$ and $y$: where
$z=L(\tau,\lambda)y$. Thus, the matrix $A_\sigma$ has the form
\begin{equation}\label{nov2010}
A_\sigma(\tau,\lambda)=L^T(\tau,\lambda)P_\sigma L(\tau,\lambda).
\end{equation}
To get an expression for the change of variables matrix
$L(\tau,\lambda)$, set $\rho_{\ell_k,j}:=\delta_{\ell_k,j}$, and
$$
\rho_{r_k,j}:=\left\{
\begin{array}{ll}
0,&\mbox{when }j>r_k,\mbox{ or }1\le j\le \ell_k\\
&\\
1,&\mbox{when }\ell_k<j\le r_k,
\end{array}
\right. .
$$
With this notation, the lower-triangular matrix $L(\tau,\lambda)$ has the form
\begin{equation}\label{nov2106}
\!L(\tau,\lambda)=\!\left[
{\begin{array}{lccccc}
\! \dfrac{\rho_{11}}{\lambda_1}\tau^{1/2}_1I_d&0&0&\ldots&0&0\\
 &&&&&\\
\dfrac{\rho_{21}}{\lambda_2}
\tau^{1/2}_1I_d&\dfrac{\rho_{22}}{\lambda_2}\tau_2^{1/2}I_d
&0&\ldots&0&0\\
% \rho_{31}\tau^{1/2}_1\lambda_3^{-1}&\rho_{32}\tau_2^{1/2}\lambda_3^{-1}
% &\rho_{33}\tau_3^{1/2}\lambda_3^{-1}&\ldots&0&0\\
\vdots&\vdots&\vdots&\vdots&\vdots&\vdots\\
\!\dfrac{\rho_{2n-1,1}}{\lambda_{2n-1}}\tau^{1/2}_1I_d
&\dfrac{\rho_{2n-1,2}}{\lambda_{2n-1}}\tau_2^{1/2}I_d
&\dfrac{\rho_{2n-1,3}}{\lambda_{2n-1}}\tau_3^{1/2}I_d
&\!\ldots&\dfrac{\rho_{2n-1,2n-1}}{\lambda_{2n-1}}\tau_{2n-1}^{1/2}I_d&0\\
 &&&&&\\
\!\dfrac{\rho_{2n,1}}{\lambda_{2n}}\tau^{1/2}_1I_d
&\dfrac{\rho_{2n,2}}{\lambda_{2n}}\tau_2^{1/2}I_d
&\dfrac{\rho_{2n,3}}{\lambda_{2n}}\tau_3^{1/2}I_d
&\ldots&\!\!\dfrac{\rho_{2n,2n-1}}{\lambda_{2n}}\tau_{2n-1}^{1/2}I_d
&\!\dfrac{\rho_{2n,2n}}{\lambda_{2n}}\tau_{2n}^{1/2}I_d
\end{array}
}
\right].
\end{equation}
The matrices $A_\sigma$, $L$ and $P_\sigma$ are all block matrices,
with $d\times d$ blocks, which are multiples of the identity matrix
$I_d$. The matrix $A_\sigma$ is  symmetric and non-negative so
$\hbox{det}(I_{2nd}+iA_\sigma)$ is the product
$\prod_{j=1}^{2nd}(1+i\mu_j)$, where $\mu_j$ are the eigenvalues of
$A_\sigma$. It is easy to see that $\mu_j$ are the eigenvalues of the
matrix $2n\times 2n$ matrix $A_\sigma^r$ obtained by reducing each $d\times d$ identity block
in $A_\sigma$ to a ``$1\times 1$'' block, except that the
corresponding multiplicities are multiplied by $d$. 
%%Just take the eigenvector of the reduced 2n\times 2n matrix and blow
%%it up to 2nd vector, and you can do d shifts in each component
%%giving multiplicty d.
%%%
We conclude that
\begin{equation}\label{nov2102}
\hbox{det}(I_{2nd}+iA_\sigma)=[\hbox{det}(I_{2n}+iA_\sigma^r)]^d.
\end{equation}
Combining the above with \eqref{051511}, we obtain 
\begin{eqnarray}
\label{051511bis}
&&c_{n,\eps}
=\left(\frac{\eps^2}{(2\pi)^{d/2}}\right)^n 
\sum_{\si}\int_1^{+\infty}\farc{d\lambda_1}{\lambda_1^{(\gamma+d)/2}}\dots
\int_1^{+\infty}\farc{d\lambda_{2n}}{\lambda_{2n}^{(\gamma+d)/2}}\prod_{j=1}^n
\delta(\lambda_{\ell_j}-\lambda_{r_j})\\
&&~~~~\times\int_{\tilde\Delta_{2n}(t/\eps^2)}
\left[{\rm det}(I_{2n}+iA_{\si}^r(\tau,\lambda))\right]^{-d/2}d\tau_{1,2n}.\nonumber
\end{eqnarray}
% \begin{equation}
% \label{051511a}
% c_{n,\eps}(i)
% =\left(\frac{\eps^2}{(2\pi)^{d/2}}\right)^n 
% \sum_{\si}\int_{\Delta_{2n}(t/\eps^2)}\left[{\rm det}(I+ia_{\si}(s))\right]^{-d/2}ds_{1,2n}.
% \end{equation}
% Denote
% $$
% \rho_{r_k,j}:=\left\{
% \begin{array}{ll}
% 0,&\mbox{when }j>r_k,\mbox{ or }1\le j\le \ell_k\\
% &\\
% 1,&\mbox{when }\ell_k<j\le r_k,
% \end{array}
% \right.
% $$
% and
% $\rho_{\ell_k,j}:=\delta_{\ell_k,j}$.
The reduced matrix $A_\sigma^r$ has the form as in (\ref{nov2010}):
$$
A_{\si}^r(\tau,\lambda)=L_r^T(\tau,\la)P_r(\si)L_r(\tau,\la),
$$
where $P_r(\si)$ is the projection on the (now scalar) $r_k$-components,
and $L_r(\tau,\la)$ has the same form (\ref{nov2106}) as $L(\tau,\lambda)$ except that each
$d\times d$ identity block is contracted to  a scalar.
% $$
% L(s)=\left[
% \begin{array}{lccccc}
% \rho_{11}\tau^{1/2}_1&0&0&\ldots&0&0\\
% \rho_{21}\tau^{1/2}_1&\rho_{22}\tau_2^{1/2}&0&\ldots&0&0\\
% \rho_{31}\tau^{1/2}_1&\rho_{32}\tau_2^{1/2}&\rho_{33}\tau_3^{1/2}&\ldots&0&0\\
% \vdots&\vdots&\vdots&\vdots&\vdots&\vdots\\
% \rho_{2n-1,1}\tau^{1/2}_1&\rho_{2n-1,2}\tau_2^{1/2}&\rho_{2n-1,3}\tau_3^{1/2}&\ldots&\rho_{2n-1,2n-1}\tau_{2n-1}^{1/2}&0\\
% \rho_{2n,1}\tau^{1/2}_1&\rho_{2n,2}\tau_2^{1/2}&\rho_{2n,3}\tau_3^{1/2}&\ldots&\rho_{2n,2n-1}\tau_{2n-1}^{1/2}&\rho_{2n,2n}\tau_{2n}^{1/2}
% \end{array}
% \right],
% $$
% recall that $\tau_k:=s_k-s_{k-1}$, with $s_0:=0$, and
% $P(\si)$ is the martix of the orthogonal projection
% onto ${\rm e}_{r_k}$, $k=1,\ldots,n$. Denote by $L_r(s)$ the $n\times
% n$ matrix obtained from $L(s)$ by removing the rows and columns that
% correspond to indices $\ell_k$, $k=1,\ldots,n$. Let
% $$
% a^{(r)}_{\si}(s):=L^T_r(s)L_r(s).
% $$
% Immediately from the above formula we obtain.
% \begin{proposition}
% \label{prop021711}
% We have
% \begin{equation}
% \label{071711}
% {\rm det}(a^{(r)}_{\si}(s))=\prod_{k=1}^n\tau_{r_k}.
% \end{equation}
% \end{proposition}
% \bigskip
Thus, Proposition~\ref{prop011711bis} is a consequence of
(\ref{nov2102}) and the following lemma.
\begin{lemma}
\label{prop031711}
For any permutation $\si$ we have
\begin{equation}
\label{111711}
|{\rm det}(I_{2n}+iA_{\si}^r(\tau,\lambda))|\ge
\frac{1}{2^{n/2}}\prod_{k=1}^n\left(1+\farc{\tau_{r_k}}{\lambda_{r_k}^2}
\right)
\end{equation}
for all $n\ge1$, $( \tau_1,\ldots,\tau_{2n})\in(0,+\infty)^{2n}$ and $(\la_{r_1},\ldots,\la_{r_n})\in(0,+\infty)^n$.
\end{lemma}
% \begin{proposition}
% \label{prop011711}
% For any permutation $\si$ we have
% \begin{equation}
% \label{011711}
% |{\rm det}(I+ia_{\si}(s))|\ge \frac{1}{2^{n/2}}\left(1+\prod_{k=1}^n\tau_{r_k}\right),
% \end{equation}
% where $(r_k)$ is the sequence of the right vertices of the bonds of 
% pairing ${\mathfrak f}(\si)$, see \eqref{021711}.
% \end{proposition}
{\bf Proof.}
The non-negative symmetric $2n\times 2n$ matrix $A_{\si}^r(\tau,\lambda)$ has eigenvalues
\[
\ga_1\ge\ga_2\ge \ga_n>\gamma_{n+1}=\ldots =\ga_{2n}=0.
\] 
In order to deal with the non-degenerate part, let us denote by
$N_r(\tau,\la)$ the $n\times n$ matrix obtained from $L_r(\tau,\la)$ 
by removing the
rows and columns that correspond to the indices $\ell_k$, 
with~$k=1,\ldots,n$. We will also consider the $n\times n$ matrix
$$
a^{(r)}_{\si}(\tau,\lambda):=N^T_r(\tau,\lambda)N_r(\tau,\lambda).
$$
Let $\mu_1\ge\mu_2\ge \ldots\ge \mu_{n}\ge0$ be the eigenvalues
of $a^{(r)}_{\si}(s)$. We claim
that
\begin{equation}
\label{061711}
\ga_{j}\ge \mu_j,\quad j=1,\ldots,n.
\end{equation}
% and
% \begin{equation}
% \label{061711a}
% \mu_j =\tau_j,\quad j=1,\ldots,n.
% \end{equation}
Consider the quadratic forms
$Q(\cdot )$ and $P(\cdot)$ on  $\mathbb R^{2n}$ and $\bbR^n$,
respectively, that correspond to the matrices $A_{\si}^r(\tau,\lambda)$ and
$a^{(r)}_{\si}(\tau,\lambda)$.
Let $H_n:=\hbox{span}(e_{r_j},j=1,\dots,n)\subset\mathbb R^{2n}$, 
and $U:\bbR^n\to H_n$ be given by
$$
Uy=\sum_{j=1}^ny_j{ e}_{r_j},
$$  
so that $Q(Uy)=P(y)$. Note that
$$
\gamma_1=\sup_{x\in\Rm^{2n},\|x\|=1}Q(x)\ge \sup_{x\in H_n,\|x\|=1}Q(x)=\sup_{y\in \bbR^n,\|y\|=1}P(y)=\mu_1.
$$
Similarly, for $1<k\le n$, let ${\cal H}_k$ the family
of all subspaces of $\mathbb R^{2n}$ of dimension $k$, ${\cal H}_k'$
the family of all
$k$-dimensional subspaces of $H_n$, and ${\cal H}_k''$ the family of
all $k$-dimensional subspaces of $\Rm^n$. 
Then, by Fisher's principle, see part (i) of Theorem 4, p. 318 of
\cite{lax},
we have
$$
\ga_{k}=\sup_{H\in {\cal H}_k}\inf_{x\in H,\|x\|=1}Q(x)\ge \sup_{H\in
  {\cal H}_k'}\inf_{x\in H,\|x\|=1}Q(x)=\sup_{H\in {\cal H}_n''}\inf_{x\in H,\|x\|=1}P(x)=\mu_k.
$$
We see that \eqref{061711} holds.
% Suppose that $f_1$ is the eigenvector of $a^{(r)}_{\si}(s)$ that
% corresponds to $\mu_1$ and 
% $K_{n-1}$ is its orthogonal complement, then
% $$
% \inf_{y\in K_{n-1},\|y\|=1}P(y)=\mu_2,
% $$
% thus
% $$
% \inf_{x\in H_{n-1},|x|=1}\Phi(x)=\inf_{y\in K_{n-1}}\Psi(y)=\mu_2,~~~H_{n-1}=U(K_{n-1}).
% $$
% It follows, again by Fisher's principle, that
% $$
% \ga_{n+2}=\sup_{H\in {\cal H}_{n-1}}\inf_{x\in H,|x|=1}\Phi(x)\ge \mu_2.
% $$
% Continuing this procedure we see that \eqref{061711} holds.
This argument allows us to write
\begin{eqnarray}
\label{091711}
&&|{\rm det}(I+iA_{\si}^r(\tau,\lambda))|=\prod_{k=1}^{2n}|(1+i\ga_k)|\ge
\prod_{k=1}^{n}|(1+i\ga_{n+k})|\ge \prod_{k=1}^{n}|(1+i\mu_{k})|\ge
% &&
% =\prod_{k=1}^{n}(1+\mu^2_{k})^{1/2}\ge
\prod_{k=1}^{n}\frac{1+\mu_{k}}{\sqrt{2}}\nonumber\\
&&=\frac{1}{2^{n/2}}\Big(1+\sum_{k=1}^n\sum_{1\le i_1,i_2,\dots,i_k\le
  n}\mu_{i_1}\mu_{i_2}\dots\mu_{i_k}
\Big).
%\ge
%\frac{1}{2^{n/2}}\left(1+\prod_{k=1}^{n}\mu_{k}\right)=\frac{1}{2^{n/2}}
%\left[1+{\rm det}(a^{(r)}_{\si}(s))\right].\nonumber
\end{eqnarray}
%Formula \eqref{011711} then follows from \eqref{071711}.\qed
In order to re-write the summation in the right side, we use an
elementary linear algebra result (see p. 88 of
\cite{gelfand}).  Recall that a $k\times k$ matrix $b$ is a principal
minor of rank $k\in\{1,\ldots,n\}$ of an~$n\times n$ matrix $B$, if it is obtained
by removing $n-k$ different rows and columns containing the diagonal
elements $b_{j_1,j_1},\ldots, b_{j_{n-k},j_{n-k}}$ for some
$j_1<j_2<\ldots <j_{n-k}$.  Then, we have
$$
\sum_{b\in {\cal M}_k(B)}{\rm det}(b)=\sum_{1\le j_1<j_2<\ldots
  <j_{k}\le n}\prod_{\ell=1}^{k}\eta_{j_\ell}.
$$
Here, $\eta_j$, $j=1,\dots,n$ are the eigenvalues of the matrix $B$,
and ${\cal M}_k(B)$ is the collection of all $k\times k$ principal
minors of the matrix $B$. Thus, (\ref{091711}) can be written as 
\begin{eqnarray}
\label{091711bis4}
&&|{\rm det}(I+iA_{\si}^r(\tau,\lambda))\ge
\frac{1}{2^{n/2}}\Big(1+\sum_{b\in {\cal M}_k(B)}{\rm det}(b)
\Big).
%\ge
%\frac{1}{2^{n/2}}\left(1+\prod_{k=1}^{n}\mu_{k}\right)=\frac{1}{2^{n/2}}
%\left[1+{\rm det}(a^{(r)}_{\si}(s))\right].\nonumber
\end{eqnarray}

In order to estimate the right side we will use the following lemma.
\begin{lemma}
\label{lm012511}
Let $b\in {\cal M}_k\big(a_{\si}^{(r)}(\tau,\lambda)\big)$ be the
principal minor obtained from $a_{\si}^{(r)}(\tau,\lambda)$ by the
removal of the rows and columns that correspond to the indices $1\le
j_1<j_2<\ldots <j_{n-k}\le n$, then
\begin{equation}
\label{101711}
{\rm det}(b)\ge\prod_{j\not\in\{j_1,\ldots,j_{n-k}\} }\frac{\tau_{r_j}}{\lambda_{r_j}^2}.
\end{equation}
 \end{lemma}
{\bf Proof.}
Let 
$\tilde l$ be the principal minor obtained from $N_r(\tau,\lambda)$ by
the removal of the rows and columns that correspond to the indices $1\le
j_1<j_2<\ldots <j_{n-k}\le n$, and
$P$ be the projection matrix onto ${\rm span}\{{\rm
  e}_{r_j},\,j\not\in\{j_1,\ldots,j_{n-k}\}\}$, then 
\[
\hbox{det}(\tilde l)=\hbox{det}(PN_rP+I-P),
\]
and
\[
\hbox{det}(\tilde l\tilde l^T)=\hbox{det}(PN_rPPN_r^TP+I-P).
\]
It follows that
\[
\hbox{det}
(b)=\hbox{det}(PN_rN_r^TP+I-P)\ge\hbox{det}(PN_rPPN_r^TP+I-P)=\hbox{det}(\tilde
l\tilde
l^T),
\] 
as seen by the comparison of the corresponding quadratic forms. 
We 
conclude that 
$$
{\rm det}(b)\ge {\rm det}(\tilde l^t\tilde l)=\prod_{j\not\in\{j_1,\ldots,j_{n-k}\} }\frac{\tau_{r_j}}{\lambda_{r_j}^2},
$$
finishing the proof of Lemma~\ref{lm012511}.
$\Box$

% *********************DO NOT REMOVE: ALTERNATIVE PROOF ************************

% Note that $b$ is the $k\times k$ matrix
% of the quadratic form
% $$
% \Phi_b(y):=(a_{\si}^{(r)}Uy,Uy)_{\bbR^n}=(U^ta_{\si}^{(r)}Uy,y)_{\bbR^k}~~y\in\Rm^k,
% $$ 
% where   $U:\bbR^k\to {\rm span}\{{\rm e}_{r_j},\,j\not\in\{j_1,\ldots,j_{n-k}\}\}$
% is given by
% $$
% Uy=\sum_{j=1}^ky_j{\rm e}_{r_j}.
% $$
% Denote also by $P$ the orthogonal projection of $\bbR^n$ onto $ {\rm
%   span}\{{\rm e}_{r_j},\,j\not\in\{j_1,\ldots,j_{n-k}\}\}$ and
% let 
% $\tilde l$ be the principal minor obtained from $N_r(\tau,\lambda)$ by
% the removal of the rows and columns that correspond to the indices $1\le
% j_1<j_2<\ldots <j_{n-k}\le n$.
% We write
% \begin{eqnarray}
% \label{012511bis}
% &&
%  \Phi_b(y)= (a_{\si}^{(r)}Uy,Uy)_{\bbR^n}=(N_r^T(\tau,\la)N_r(\tau,\la)Uy,Uy)_{\bbR^n}
%  \nonumber\\
%  &&
%  \\
%  &&
% \ge 
%  (PN_r(\tau,\la)Uy,PN_r(\tau,\la)Uy)_{\bbR^n} =(\tilde l y,\tilde ly)_{\bbR^d}=:\Psi(y),\quad y\in\bbR^k.\nonumber
% \end{eqnarray}
% Thanks to the fact that the comparison of the forms implies an
% analogous comparison of the determinants of their respective matrices we 
% conclude that 
% $$
% {\rm det}(b)\ge {\rm det}(\tilde l^t\tilde l)=\prod_{j\not\in\{j_1,\ldots,j_{n-k}\} }\frac{\tau_{r_j}}{\lambda_{r_j}^2}.
% $$
% \qed

% *********************END OF  ALTERNATIVE PROOF ************************

Using \eqref{101711} in \eqref{091711} we conclude that
\begin{eqnarray}
\label{111711bis}
|{\rm det}(I+iA_{\si}^r(\tau,\lambda))|\ge \frac{1}{2^{n/2}}
\prod_{k=1}^n\left(1+\farc{\tau_{r_k}}{\lambda_{r_k}^2}\right).
\end{eqnarray}
This finishes the proof of Lemma~\ref{prop031711}, and thus also that
of Proposition~\ref{prop011711bis}. $\Box$

\subsubsection*{The case $\xi\neq 0$}

As for $\xi=0$, we start with (\ref{031911}) also for $\xi\neq 0$, except
now we have to take into account the contribution of 
the matrix $C_\si(\tau,\lambda)$.
The matrix $A_\si(\tau,\lambda)$ is symmetric and non-negative, 
hence~$C_\si(\tau,\lambda)$ is  diagonalizable with respect to the orthonormal
basis of eigenvectors of $A_\si(\tau,\lambda)$: 
\begin{equation}\label{nov3004}
A_\si(\tau,\lambda)f_j=\gamma_jf_j,~~~\gamma_j\ge 0,
\end{equation}
and
\[
C_\si(\tau,\lambda)f_j=\mu_jf_j,~~\mu_j=\frac{\gamma_j}{1+i\gamma_j},
\]
% Suppose that $(f_j)$ is an orthonormal base that diagonalizes $A_{\si}(s)$ and that
% $$
% A_{\si}(s)f_j=\ga_j f_j,\quad j=1,\ldots,2nd
% $$
% where $\ga_1\ge \ldots\ge\ga_{2nd}\ge 0$.
so that
\begin{equation}\label{nov3002}
(C_{\si}(\tau,\lambda)\Xi,\Xi)_{\bbR^{2nd}}=\sum_{j=1}^{2nd}\frac{\ga_j}{1+i\ga_j}(\Xi,f_j)^2_{\bbR^{2nd}},
\end{equation}
thus
\begin{equation}
\label{020412}
\left|\exp\left\{-\frac{1}{2}(C_{\si}(\tau,\lambda)\Xi,\Xi)_{\bbR^{2nd}}\right\}\right|=\prod_{j=1}^{2nd}\exp\left\{-\frac{\ga_j}{2(1+\ga_j^2)}(\Xi,f_j)^2_{\bbR^{2nd}}\right\}
\le 1.
\end{equation}
As a consequence, we have an estimate
\begin{equation}
\label{041911}
|c_{n,\eps}(t,\xi)|
\le  \left(\frac{\eps^2}{(2\pi)^{d/2}}\right)^n
\sum_{\si}\int_1^{+\infty}\frac{d\la_{r_1}}{\la_{r_1}^{\ga+d}}\ldots \int_1^{+\infty}\frac{d\la_{r_n}}{\la_{r_n}^{\ga+d}}\int_{\tilde\Delta_{2n}(t/\eps^2)}\left|{\rm det}[I_{2nd}+i A_{\si}(\tau,\lambda)]\right|^{-1/2}d\tau_{1,2n}
\end{equation}
From this point on, we can repeat verbatim the estimates
for $\xi=0$, and the conclusion of Proposition ~\ref{prop1811}
can be extended to the case $\xi\not=0$ as well. 

\subsection*{Proof of Theorem \ref{thm-may2302}}

Proposition~\ref{prop1811} allows us to pass to the limit $\eps\to 0$ termwise
in the expression \eqref{012411} for $c_{n,\eps}(t,\xi)$, so that
\begin{equation}
\label{012411a}
\lim_{\eps\to0+}\bar \zeta_\eps(t,\xi)=\hat \psi_0(\xi)\sum_{n=0}^{+\infty}\frac{(-1)^n}{(2n)!!}\lim_{\eps\to0+}c_{n,\eps}(t,\xi),
\end{equation}
provided that $t\in[0,t_0]$, where $t_0$ is so small that $Ct_0<1$,
with $C$ as in \eqref{121711a}. 
We will again assume that $s(\lambda)=1$ to simplify the notation.
Let us go back to representation~(\ref{031911}):  
\begin{eqnarray}
\label{031911a}
&&
c_{n,\eps}(t,\xi)
=\left(\frac{\eps^2}{(2\pi)^{d/2}}\right)^n \sum_{\si}\int_1^{+\infty}\frac{d\la_{r_1}}{\la_{r_1}^{\ga+d}}\ldots \int_1^{+\infty}\frac{d\la_{r_n}}{\la_{r_n}^{\ga+d}}\int_{\tilde\Delta_{2n}(t/\eps^2)}d\tau_{1,2n}\\
&&
~~~~~~~~~~\times{\rm det}( I+iA_{\si}(\tau,\lambda))^{-1/2}\exp
\left\{-\frac{1}{2}(C_{\si}(\tau,\lambda)\Xi(\tau),\Xi(\tau))_{\bbR^{2nd}}\right\}.\nonumber
\end{eqnarray}
Each term appearing in the sum
in the right side of  \eqref{031911a} is of the form
\begin{equation}
\label{031911c}
\frac{1}{(2\pi)^{nd/2}}\int_1^{+\infty}\!\!\!\ldots
\int_1^{+\infty}\prod_{j=1}^n\la_{r_j}^{-\ga-d} d\la_{r_1,r_n}\int_{\tilde{\Delta}_{n}(t)}\Theta_\eps(\tau_\eps;\si)d\tau_{\ell_1,\ell_n}.
\end{equation}
Here $d\la_{r_1,r_n}=d\la_{r_1}\ldots d\la_{r_n}$,  $d\tau_{\ell_1,\ell_n}=d\tau_{\ell_1}\ldots d\tau_{\ell_n}$ and  the domain of integration is
$$
\tilde{\Delta}_{n}(t;\si):=\Big[(\tau_{\ell_1},\ldots,\tau_{\ell_n}):\,\sum_{j=1}^n\tau_{\ell_j}\le
t,\,\tau_{\ell_j}\ge0,\,j=1,\ldots,n\Big].
$$
The integrand in (\ref{031911c}) is defined as follows:
set $\tau_\eps:= (\tau_{1,\eps},\ldots,\tau_{2n,\eps})$, with
$\tau_{\eps,\ell_j}:=\eps^{-2}\tau_{\ell_j}$,  and~$\tau_{\eps,r_j}':=\tau_{r_j}$.   
Then
\begin{equation}
\label{072811}
\Theta_\eps(\tau_\eps;\si):=\int_{\widehat{\Delta}_{n}((t-\tau)/\eps^2)}{\rm det}( I+iA_\si(\tau_\eps,\la))^{-1/2}\exp
\left\{-\frac{1}{2}(C_{\si}(\tau_\eps,\lambda)\Xi(\tau_\eps),\Xi(\tau_\eps))_{\bbR^{2nd}}\right\}d\tau_{r_1,r_n},
\end{equation}
where $\tau:=\sum_{j=1}^n\tau_{\ell_j}$, $d\tau_{r_1,r_n}:=d\tau_{r_1}\ldots d\tau_{r_n}$ and 
$$
\widehat{\Delta}_{n}(u;\si):=\left[(\tau_{r_1},\ldots,\tau_{r_n}):\,\sum_{j=1}^n\tau_{r_j}\le
u,\,\tau_{r_j}\ge0,\,j=1,\ldots,n\right].
$$
%Let $\mathbb C_-:=\mathbb C\setminus(i(-\infty,0])$. 
We will distinguish in the computation of the limit
between simple and non-simple pairings -- note that no such distinction was made
in the estimates so far.

\subsubsection*{Non-simple pairings}
Recall that the
pairing ${\mathfrak e}:=\{(1,2),(3,4),\ldots,(2n-1,2n)\}$ is called {\em simple}.
\begin{lemma}
\label{lm010112}
For any  
$(\la_{r_1},\ldots,\la_{r_n})\in(1,+\infty)^{n}$ and
$(\tau_1,\ldots,\tau_{2n})\in(0,+\infty)^{2n}$, we have
\begin{equation}
\label{010112}
\lim_{\eps\to0+}|{\rm det}( I+iA_\si(\tau_\eps,\la))|=+\infty,
\end{equation}
provided that $\si$ is a permutation such that ${\mathfrak
  f}(\si)\not={\mathfrak e}$ (see \eqref{052811} for the definition of
the map
${\mathfrak f}$).
\end{lemma}
{\bf Proof.} 
Note that if $a_{\ell,\ell}\not=0$ for some left vertex $\ell$
of ${\mathfrak f}(\si)$
then choosing $y=(y_1,\ldots,y_{2n})$ with~$y_j=0$ for $j\not=\ell$
and $y_{\ell}=e$ for some $e\in\bbR^d$ such that $|e|=1$,
we get, using \eqref{Phi} and~\eqref{A}:
$$
%\lim_{\eps\to0+}\Phi_{\si}(y)=
\lim_{\eps\to0+}\left(A_{\si}(\tau_\eps,\lambda)y,y\right)_{\bbR^{2dn}}
=\lim_{\eps\to0+} a_{\ell,\ell}\frac{\tau_\ell}{\eps^2}=+\infty.
$$
It follows that the largest eigenvalue of $A_\sigma$ satisfies
$\ga_{2n}\to+\infty$, 
as $\eps\to0+$, and~\eqref{010112} follows. On the other hand, if $\si$ is such that 
\begin{equation}
\label{020112}
a_{\ell,\ell}=0\quad\mbox{ for all left vertices $\ell$  of ${\mathfrak f}(\si)$}
\end{equation}
then, according to the defintion~\eqref{aj} of $a_{\ell,\ell}$, 
for any left vertex $\ell$ of  ${\mathfrak f}(\si)$ there is no  
bond~$(\ell',r')$ such that $\ell'<\ell<r'$. 
This implies that for all bonds we have
$\ell=r-1$. Indeed, otherwise we would let $\ell$ be the smallest left
vertex for which $r\neq \ell+1$. Then $\ell+1$ would have to be a left
vertex for which $a_{\ell+1,\ell+1}\neq 0$, giving a contradiction to~\eqref{020112}. 
% if we have $\ell <r-1$ and all $r'\in (\ell, r)$
% were
% right ones then for such a vertex $r'$ its left counterpart $\ell'$
% would satisfy $\ell'<\ell<r'$. Then however we would have
% $a_{\ell',\ell'}>0$, which would contradict \eqref{020112}. 
This
proves that ${\mathfrak f}(\si)={\mathfrak e}$.
\qed

Since, according to \eqref{011711bis}, there exists a constant $C>0$ such that
\begin{equation}
\label{040112}
|{\rm det}( I+iA_\si(\tau_\eps,\la))|^{-1/2}\le
C\prod_{k=1}^n\left(1+\farc{\tau_{r_k}}{\lambda_{r_k}^2}
\right)^{-d/2},
\end{equation} 
for all permutations $\si\in\Pi(2n)$, $(\tau_1,\ldots,\tau_{2n})$,
and $(\la_{r_1},\ldots,\la_{r_n})$, we conclude by
the Lebesgue dominated convergence theorem that
$$
\lim_{\eps\to0+}\Theta_\eps(\tau_\eps;\si)=0,
$$
provided that ${\mathfrak f}(\si)\not={\mathfrak e}$. Using the same
theorem once again in (\ref{031911c}), we conclude that the limit as
$\eps\to 0^+$ of the terms in (\ref{031911a}) corresponding to such
permutations, vanishes.

\subsubsection*{Simple pairings}

Observe that for any $\si$ such that ${\mathfrak f}(\si)={\mathfrak
  e}$, we have $a_{mj}=0$ if $m\neq j$ and $a_{\ell,\ell}=0$ if $\ell$
is a left vertex, in other words, if $\ell$ is odd. The matrix
$A_\sigma$ has then a particularly simple form
\begin{equation}
\label{A1}
A_{\si}(\tau,\la)=\left[
\begin{array}{lcccccc}
0&0&0&\vdots&0&0\\
0&I_d\la_2^{-2}\tau_2&0&\vdots&0&0\\
0&0&0&\vdots&0&0\\
\vdots&\vdots&\vdots&\vdots&\vdots&\vdots\\
0&0&0&\vdots&0&0\\
0&0&0&\vdots&0&I_d\la_{2n}^{-2}\tau_{2n}
\end{array}
\right]
\end{equation}
and the matrix $C_\sigma$ has the form
\begin{equation}
\label{C1}
C_{\si}(\tau,\la)=\left[
\begin{array}{lccccc}
0&0&0&\vdots&0&0\\
0&I_d(\la_{2}^2+i\tau_{2})^{-1}\tau_2&0&\vdots&0&0\\
0&0&0&\vdots&0&0\\
\vdots&\vdots&\vdots&\vdots&\vdots&\vdots\\
0&0&0&\vdots&0&0\\
0&0&0&\vdots&0&I_d(\la_{2n}^2+i\tau_{2n})^{-1}\tau_{2n}
\end{array}
\right].
\end{equation}
Thus, we obtain for such $\sigma$:
\begin{equation}
\label{072811a}
\lim_{\eps\to0+}\Theta_\eps(\tau_\eps;\si)=\prod_{k=1}^{n}\int_0^{+\infty}
\frac{1}{(1+i\la_{2k}^{-2}\tau)^{d/2}}\exp
\left\{-\frac{|\xi|^2\tau^2}{2(\la_{2k}^2+i\tau)}\right\}d\tau.
\end{equation}
Since, as we already
know, there are precisely $(2n)!!$ permutations that yield ${\mathfrak
 f}(\si)\not={\mathfrak e}$ we conclude from \eqref{031911a}-\eqref{031911c}and (\ref{072811a})
  that
\begin{equation}
\label{c-n2bis}
\lim_{\eps\to0+}c_{n,\eps}(\xi,t)=\left[ {-2itr(\xi)} \right]^n
%\left[\frac{2tr(\xi)}{(2\pi)^{d/2}}\right]^n, 
\end{equation}
with
\begin{equation}
\label{050112}
r(\xi)=\farc{i}{2}\farc{1}{(2\pi)^{d/2}}\int_1^{+\infty}\frac{\kappa(\lambda,\tau)d\la}{\la^{\ga+d-2}},~~~
\kappa(\lambda,\tau)=\int_0^{+\infty}\frac{1}{(1+i\tau)^{d/2}}\exp
\left\{-\frac{(\la|\xi|\tau)^2}{2(1+i\tau)}\right\}d\tau.
\end{equation}
%\begin{equation}
%\label{050112}
%r(\xi)=\int_1^{+\infty}\frac{d\la}{\la^{\ga+d-2}}\int_0^{+\infty}\frac{1}{(1+i\tau)^{d/2}}\exp
%\left\{-\frac{(\la|\xi|\tau)^2}{2(1+i\tau)}\right\}d\tau.
%\end{equation}
This leads to the conclusion of the theorem. $\Box$

\subsection*{Proof of Theorem \ref{main1}}

In the setting of Theorem~\ref{main1}, we conclude that there exists $t_0>0$ such that
\begin{equation}
\label{012411c1}
\bar \zeta_\eps(t,\xi)=\hat
\psi_0(\xi)\sum_{n=0}^{+\infty}\frac{(-1)^n}{(2n)!!}\lim_{\eps\to0+}c_{n,\eps}(t,\eps^{\beta}
\xi),
\end{equation}
with $c_{n,\eps}(t,
\xi)$ given by \eqref{031911a} and $t\in[0,t_0]$. 
The same computations as in the proof of Theorem~\ref{thm-may2302}
show that
$$
\lim_{\eps\to0+}c_{n,\eps}(t,\eps^{\beta}\xi)=\left[-2itr(0)\right]^n ,\quad n\ge0,\,t\ge0.
%\left[\frac{2\kappa(0)t}{(2\pi)^{d/2}}\right]^n ,\quad n\ge0,\,t\ge0.
$$ 
%\textcolor{blue}{\bf How do we know that the norm $\|\hat
%  \zeta_\eps(t,\cdot)\|_{L^2(\bbR^d)}$ is constant in time? Can we
%  cite some source? Note that we are dealing here with $V(x)$ that is
%  neither integrable, nor bounded.}
We conclude that for any $t\in[0,t_0]$
\begin{equation}
\label{012411c2a}
\lim_{\eps\to0+}\bar \zeta_\eps(t,\xi)=\bar \zeta(t,\xi),
\end{equation}
with $\bar\zeta(t,\xi)$ given by \eqref{may2338}, both pointwise in $\xi$
and weakly in $L^2(\bbR^d)$. In order to show that not only the limit holds   
for the expectation, but actually the limit is deterministic, 
observe that, due to \eqref{011708}, we have
\begin{eqnarray}
\label{010312}
&&
\bbE\|\hat \zeta_\eps(t,\cdot)-\bar
\zeta(t,\cdot)\|_{L^2(\bbR^d)}^2=\bbE\|\hat
\zeta_\eps(t,\cdot)\|^2+\|\bar
\zeta(t,\cdot)\|^2-
2{\rm Re}\left(\bar \zeta_\eps(t,\cdot),\bar
  \zeta(t,\cdot)\right)_{L^2(\bbR^d)}\nonumber\\
&&
=2\|\hat
\psi_0\|^2-2{\rm Re}\left(\bar \zeta_\eps(t,\cdot),\bar
  \zeta(t,\cdot)\right)_{L^2(\bbR^d)}.
\end{eqnarray}
Letting $\eps\to0+$ and using \eqref{012411c2a} we conclude
that the right hand side of \eqref{010312} tends to 
$$
2\|\hat
\psi_0\|^2-2\|\bar \zeta(t,\cdot)\|_{L^2(\bbR^d)}=0,
$$
which ends the proof of the theorem.
$\Box$

\section{The infinite effective potential regime}

\label{sec5.3.2}

In the present section we give the proof of Theorem \ref{main1a}.
%
%{\bf Bounds on $c_{n,\eps}(t,\xi)$.} 
%
Adjusting for the scaling, we can write an analogue of \eqref{041911},
though this time the simplex  $\tilde\Delta_{2n}(t/\eps^2)$ is replaced by
$\tilde\Delta_{2n}(t/\eps^{2\al})$. Using Proposition
\ref{prop011711bis}, we arrive at the estimate:
\begin{equation}
\label{051511cbis1}
|c_{n,\eps}(t,\xi)|
\le C^n\eps^{2n} 
\sum_{\si}\int_{\tilde\Delta_n(t/\eps^{2\al})}d\tau_{\ell_1,\ell_n}
 \prod_{k=1}^n \left[\int_1^{+\infty}\frac{d\la_{r_k}}{\la_{r_k}^{\ga+d}}\int_0^{t\eps^{-2\al}}\frac{d\tau_{r_k}}{(1+\la_{r_k}^{-2}\tau_{r_k})^{d/2}}\right].
\end{equation}
Changing variables 
%$\tau_{\ell_k}':=\eps^{2\al}\tau_{\ell_k}$ and 
$\tau_{r_k}':=\la_{r_k}^{-2}\tau_{r_k}$ we obtain
\begin{eqnarray}
\label{051511cbis3}
&&
|c_{n,\eps}(t,\xi)|
\le C^n\eps^{2n} 
\sum_{\si}\prod_{k=1}^n\left[\int_1^{+\infty}\frac{d\la_{r_k}}{\la_{r_k}^{\ga+d-2}}\phi\left(\frac{t}{\la^2_{r_k}\eps^{2\al}}\right)\right]\int_{\tilde\Delta_n(t/\eps^{2\al})}d\tau_{\ell_1,\ell_n}\nonumber\\
  &&
  ~~~~~~ ~~~~~~ \le
     (2n)!\farc{(Ct\eps^{2-2\al})^n}{n!}\left[\int_1^{+\infty}\phi\left(\frac{t}{\la^2\eps^{2\al}}\right)\frac{d\la}{\la^{\ga+d-2}}\right]^n,
\end{eqnarray}
where
$$
\phi(u):=\int_0^{u}\frac{d\tau}{(1+\tau)^{d/2}}.
$$
Note that 
\begin{equation}
\label{020912}
\lim_{u\to+\infty}\phi(u)=\frac{2}{d-2},\quad d\ge 3.
\end{equation}
We change variables 
$$
u:=\frac{t}{\la^2 \eps^{2\al}},\quad k=1,\ldots,n,
$$
in the  integral appearing in the right   side of  \eqref{051511cbis3}. Taking into account \eqref{010912}, the right 
side of (\ref{051511cbis3}) can be then
rewritten in the form
\begin{equation}
\label{030912}
 (2n)!\farc{(Ct^H)^n}{n!} \left[\int_0^{t/\eps^{\al}}\phi\left(u\right)u^{(\ga+d-5)/2}du\right]^n,
\end{equation}
for 
\begin{equation}
\label{hurst}
H:=\frac{5-d-\ga}{2}
\end{equation}
and an appropriate constant $C>0$. Since $(\ga+d-5)/2<-1$ the integral  in \eqref{030912}
converges at $+\infty$, due to \eqref{020912}. As $\phi(u)\sim u$, for $u\ll1$ we conclude that it is also convergent
close to $0$, as $(\ga+d-3)/2>-1$.
We conclude therefore that there exists $C>0$ such that
\begin{equation}
\label{051511cbis4}
|c_{n,\eps}(t,\xi)|
\le (2n)!\farc{(Ct^H)^n}{n!}
\left[\int_0^{t/\eps^{\al}}\phi\left(u\right)u^{(\ga+d-5)/2}du\right]^n,\quad n\ge0,\,t>0,\,\eps>0.
\end{equation}

\subsubsection*
{Computation of the limit} 

Having the uniform bound (\ref{051511cbis4}), we now compute the limit.
From \eqref{031911}, we obtain
\begin{eqnarray}
\label{040912}
&&
c_{n,\eps}(t,\xi)
= \left(\frac{\eps^2}{(2\pi)^{d/2}}\right)^n
\sum_{\si}\int_1^{+\infty}\frac{d\la_{r_1}}{\la_{r_1}^{\ga+d}}\ldots \int_1^{+\infty}\frac{d\la_{r_n}}{\la_{r_n}^{\ga+d}}\int_{\Delta_{2n}(t/\eps^{2\al})}d\tau_{1,2n}\\
&&
~~~~~~~~~~~\times{\rm det}( I+iA^r_{\si}(\tau,\lambda))^{-d/2}\exp
\left\{-\frac{\eps^{2\beta}|\xi|^2}{2}(C_{\si}^r(\tau,\la)\top^{1/2}{\bf 1}\cdot \top^{1/2}{\bf 1}) \right\},\nonumber
\end{eqnarray}
Here, we denote
$
\top:={\rm diag}[\tau_1,\ldots,\tau_{2n}],
$ 
$
{\bf 1}^T:=\underbrace{[1,\ldots,1]}_{2n}
$
and
$$
C_{\si}^r(\tau,\la):=A_{\si}^r(\tau,\la)[I_{2n}+iA_{\si}^r(\tau,\la)]^{-1}.
$$
We change variables $\tau_k':=\eps^{2\al}\tau_k/t$,  $k=1,\ldots,2n$, $u_k:=t\la_{r_k}^{-2}\eps^{-2\al}$, $k=1,\ldots,n$,
and recall that
\[
H=\farc{1}{\alpha}=\farc{5-\gamma-d}{2}.
\]
We can write then
\begin{eqnarray}
\label{040912a}
&&
c_{n,\eps}(t,\xi)
= \left(\frac{t^{H}}{2(2\pi)^{d/2}}\right)^n
\sum_{\si}\int_0^{t/\eps^{2\al}}u_1^{(\ga+d-3)/2}du_1\ldots \int_0^{t/\eps^{2\al}}u_n^{(\ga+d-3)/2}du_n\int_{\tilde\Delta_{2n}(1)}d\tau_{1,2n}\nonumber
\\
&&
~~~~~~~~~~\times{\rm det}( I+i\tilde A^r_{\si}(\tau,u))^{-d/2}\exp
\left\{-\frac{t\eps^{2(\beta-\al)}|\xi|^2}{2}(\tilde C_{\si}^r(\tau,u)\top^{1/2}{\bf 1},\top^{1/2}{\bf 1})_{\bbR^{2nd}}\right\}, 
\end{eqnarray}
where
$\tilde A_{\si}^r(\tau,u)=[\tilde a_{j,m}]$ is an $2n\times 2n$ matrix
satisfying, as in \eqref{aj}--\eqref{aj2}:
$$
\tilde a_{1,j}=\tilde a_{j,1}=0\quad\mbox{ for all }j\in\{1,\ldots,2n\},
$$ 
and  
\begin{equation}
\label{taj}
\tilde a_{j,j}:=\tau_j\sum_k\!\vphantom{1}'u_k,~~\hbox{for each $j\in\{2,\ldots,2n\}$}. 
\end{equation}
The summation above extends over those $k$-s, for which
$
\ell_k< j\le r_k.
$
Given $m<j$ we let
$$
\tilde a_{m,j}=\tilde a_{j,m}=(\tau_m\tau_j)^{1/2}\sum_k\!\vphantom{1}'u_k
$$
and the summation extends over  those $k$-s, for which
$
\ell_k< m< j\le r_k.
$
We also let 
$$
\tilde C_{\si}^r(\tau,u):=\tilde A_{\si}^r(\tau,u)(I_{2n}+i\tilde A_{\si}^r(\tau,u))^{-1}.
$$
 In the limit we get
\begin{eqnarray}
\label{040912b}
&&
\bar c_n(t,\xi):=\lim_{\eps\to0+}c_{n,\eps}(t,\xi)
= \left(\frac{t^{H}}{2(2\pi)^{d/2}}\right)^n
\sum_{\si}\int_0^{+\infty}u_1^{(\ga+d-3)/2}du_1\ldots \int_0^{+\infty}u_n^{(\ga+d-3)/2}du_n\nonumber\\
&&
\\
&&
~~~~~~~~~\times\int_{\tilde\Delta_{2n}(1)}d\tau_{1,2n}{\rm det}( I+i\tilde A^r_{\si}(\tau,u))^{-d/2}.\nonumber
\end{eqnarray}
Repeating the  calculations that lead to \eqref{051511} from
\eqref{011511}, this time in the reverse order, we obtain
\begin{equation}
\label{011511a}
\bar c_{n}(t,\xi)=\mathbb M\left\{\frac{t^{H}}{2(2\pi)^{d/2}}\int_0^{+\infty}u^{(\ga+d-3)/2}du
\int_0^{1}ds\int_0^{1} ds' \exp
\left\{-\frac{iu}{2}|B_s-B_{s'}|^2\right\} \right\}^n,
\end{equation}
where $B_t$ is the $d$-dimensional standard Brownian motion and
$\mathbb M$ is the corresponding expectation.
Using the bound \eqref{051511cbis4}
we conclude that there exists $t_0>0$ such that it is possible to
interchange the limit as $\eps\to0+$ with the summation over $n$, as in
\eqref{012411a} above. Performing the summation
$\sum_{n=0}^{+\infty}\bar c_{n}(t,\xi)$ we arrive at \eqref{011012}.

\bigskip

\section{Proof of Proposition \ref{prop-ser0}} \label{sec:aux}

The
conclusion of part (i) of the proposition
and formula \eqref{exp-series} follow, provided we can show that
\[
\sum_{n=0}^\infty[\Em|\hat\zeta_n(t,\xi)|^2]^{1/2}<+\infty.
\]
We have
\begin{eqnarray}\label{zeta-n1}
&&\!\!\!\!\!\!\!
\mathbb E|\hat\zeta_n(t,\xi) |^2=\frac{1}{ (2\pi)^{2nd}}
\int\limits_{\Delta_n(t)}ds_{1,n} \int\limits_{\Delta_n(t)}d\tilde{
  s}_{1,n}
\int_{\bbR^{2dn}} \mathbb E\left[\prod_{k=1}^n
\hat V(dp_k)\,\prod_{m=1}^n
\hat V^*(d\tilde p_m)\right]\\
\!
&&
\!\!\!\!\!\!\!\times\hat\psi_0(\xi-\sum_{j=1}^np_j) \hat\psi_0^*(\xi-\sum_{j=1}^n\tilde p_j) e^{iG_n}
e^{-i\tilde G_n}
%\nonumber\\
%&&
\le \farc{C^nt^{2n}\|\hat\psi_0\|_{\infty}^2}{(n!)^2} \int_{\bbR^{2dn}}  \mathbb E\left[\prod_{k=1}^n
\hat V(dp_k)\,\prod_{m=1}^n
\hat V^*(d\tilde p_m)\right]\nonumber.
\end{eqnarray}
Here $d\tilde s_{1,n}=d\tilde s_1\ldots d\tilde s_n$ and $\tilde G_n$
is given by \eqref{G-n}, with the variables $s_{1,n}, p_{1,n}$
replaced by
 $\tilde s_{1,n},\tilde p_{1,n}$.
Using the rule for the expectation of a product of $2n$ mean zero
Gaussian random variables, we may estimate the above as
\begin{eqnarray}\label{may2348}
\mathbb E|\hat\zeta_n(t,\xi) |^2\le \frac{C^nt^{2n}(2n-1)!!}{n!^2}
\left(\int_{\bbR^d}  \hat R(p)dp\right)^n\|\hat\psi_0\|_{\infty}^2
\le\frac{C^n}{n!},
\end{eqnarray}
with a constant $C>0$ independent of $n$, and part (i) of the
proposition follows.  Part (ii) is a simple consequnce of the fact that
we can interchange the summation and 
expectation in~\eqref{zeta-expand}, by virtue of the estimate
obtained in the proof of part (i). This, combined with the fact
that the odd moments vanish, yields \eqref{exp-series}.

Concerning part (iii), let us take a smooth radially symmetric
non-negative
function $\theta$ such that $\theta(x)=1$ for $|x|\le 1$ and $\theta(x)=0$
for $|x|>2$.
%$\int_{\bbR^d}\theta(x)dx=1$. 
%We shall also assume that
%$\theta(\cdot)$ is even so its Fourier transform $\hat\theta(\cdot)$
%is real valued.
To prove \eqref{011708}, we consider the regularization of
the potential: $V_R(x):=\theta_R(x)V(x)$, with
$\theta_R(x):=\theta(x/R)$. 
%The Fourier transform
%of $V_R(x)$ is
%\[
%\hat
%V_R(p)=\int_{\bbR^d}\hat\theta_R(p-q)\hat V(dq).
%\]
Let~$\hat\zeta^{(R)}(t,\xi)$ and~$\hat\zeta_n^{(R)}(t,\xi)$ be
the random fields given by modifications of formulas~\eqref{zeta-expand} 
and~\eqref{zeta-n}, with the spectral measure $\hat V(dp)$ 
replaced by $\hat V_R(p)dp$. The function  
\[
\hat \psi^{(R)}(t,\xi):=e^{-it|\xi|^2/2}\hat\zeta^{(R)}(t,\xi),\quad t\ge0,
\]
is the solution of the Schr\"odinger equation with the potential $V_R(x)$.
%Thanks to  \eqref{schoen1abis}  and the smoothness of $\theta_R$,
%given any integer $N>0$ and $R>0$
%one can find a non-negative random variable  $L$
%such that  
%\[
%\hbox{$|\hat V_R(p)|\le L(1+|p|^2)^{-N}$ for all $p\in\bbR^d$ a.s.}.
%\]
%a.s. in the realization of $\left(V(x)\right)$.
%From the series defining $\hat\zeta^{(R)}(t,\xi)$ , see \eqref{zeta-expand} ,
%we conclude easily that 
The $L^2$-norm conservation for the solutions of the Schr\"odinger equaiton
with a decaying potential means that 
\[
\|\hat \psi^{(R)}(t)\|_{L^2(\bbR^d)}=\|\hat \psi_0\|_{L^2(\bbR^d)},
\]
thus also
\[
\hbox{$\|\hat \zeta^{(R)}(t)\|_{L^2(\bbR^d)}=\|\hat
\psi_0\|_{L^2(\bbR^d)}$ for all $t,R>0$ a.s.}
\]
Note that
$$
\|\hat \psi_0\|_{L^2(\bbR^d)}^2=\bbE\|\hat \zeta^{(R)}(t)\|_{L^2(\bbR^d)}^2=
\sum_{m,n\ge0}\bbE\left[ \left(\hat\zeta^{(R)}_n(t),\hat\zeta^{(R)}_m(t)\right)_{L^2(\bbR^d)}\right].
$$
By the Cauchy-Schwarz inequality, the absolute value of the term of the
series on the
right hand side is bounded from above by $(a_{n,R}a_{m,R})^{1/2}$,
where 
$$
a_{n,R}:=\mathbb E\int_{\bbR^d}|\hat\zeta_n^{(R)}(t,\xi) |^2d\xi.
$$
Using an analogue of \eqref{zeta-n1} for $\hat V_R(p)$ we conclude
that
$$
a_{n,R}\le \frac{C^nt^{2n}(2n-1)!!}{n!^2}
\left(\int_{\bbR^d}  \hat R(p)dp\right)^n\|\hat\psi_0\|_{L^2(\bbR^d)}^2,
$$
with the constant $C>0$ independent of $R>0$ and $n\ge0$. It follows that
that 
\begin{equation}\label{011808}
\|\hat \psi_0\|_{L^2(\bbR^d)}^2=\lim_{R\to+\infty}\bbE\|\hat \zeta^{(R)}(t)\|_{L^2(\bbR^d)}^2=\sum_{m,n\ge0}\lim_{R\to+\infty}\bbE\left[ \left(\hat\zeta^{(R)}_n(t),\hat\zeta^{(R)}_m(t)\right)_{L^2(\bbR^d)}\right].
\end{equation}
It is elementary calculation to verify that
$$
\lim_{R\to+\infty}\bbE\left[ \left(\hat\zeta^{(R)}_n(t),\hat\zeta^{(R)}_m(t)\right)_{L^2(\bbR^d)}\right]=\bbE\left[ \left(\hat\zeta_n(t),\hat\zeta_m(t)\right)_{L^2(\bbR^d)}\right]
$$
for each $n,m\ge0$.
Therefore, the right hand side of \eqref{011808} equals $\bbE\|\hat \zeta(t)\|_{L^2(\bbR^d)}^2$,
and \eqref{011708} follows.

\end{document}